\documentclass[12pt,graybox]{svmult}

\usepackage[margin=2cm]{geometry}

\usepackage{makeidx}
\usepackage{graphicx}
\usepackage{multicol}
\usepackage[bottom]{footmisc}
\usepackage{newtxtext}
\usepackage{newtxmath}
\usepackage{caption}
\usepackage[caption=false]{subfig}
\usepackage{epsfig}
\usepackage{psfrag}
\usepackage{color}
\usepackage{algorithm}
\usepackage[noend]{algorithmic}
\usepackage{url}
\usepackage{cite}
\usepackage{wrapfig}
\usepackage{amsmath}
\usepackage{fancyhdr}
\usepackage{multirow}
\usepackage[utf8]{inputenc}
\usepackage{tabu} 
\usepackage{tabularx}
\usepackage{xspace}
\usepackage{makecell}

\makeindex

\newcolumntype{P}[1]{>{\centering\arraybackslash}p{#1}}

\newenvironment{myitemize}{\begin{list}{$\bullet$}{\renewcommand{\leftmargin}{0.15in}}}{\end{list}}

\begin{document}

\title*{Ambient Intelligence for Next-Generation AR}
\author{Tim Scargill, Sangjun Eom, Ying Chen, Maria Gorlatova}
\institute{Tim Scargill \at Electrical and Computer Engineering Department, Duke University, Durham, NC, USA \email{ts352@duke.edu}
\and Sangjun Eom \at Electrical and Computer Engineering Department, Duke University, Durham, NC, USA, \email{sangjun.eom@duke.edu}
\and Ying Chen \at Electrical and Computer Engineering Department, Duke University, Durham, NC, USA, \email{ying.chen151@duke.edu}
\and Maria Gorlatova \at Electrical and Computer Engineering Department, Duke University, Durham, NC, USA, \email{maria.gorlatova@duke.edu}}
%
%
\maketitle


\abstract{Next-generation augmented reality (AR) promises a high degree of \emph{context-awareness} -- a detailed knowledge of the environmental, user, social and system conditions in which an AR experience takes place. This will facilitate both the closer integration of the real and virtual worlds, and the provision of context-specific content or adaptations. However, environmental awareness in particular is challenging to achieve using AR devices alone; not only are these mobile devices’ view of an environment spatially and temporally limited, but the data obtained by onboard sensors is frequently inaccurate and incomplete. This, combined with the fact that many aspects of core AR functionality and user experiences are impacted by properties of the real environment, motivates the use of \emph{ambient IoT devices}, wireless sensors and actuators placed in the surrounding environment, for the measurement and optimization of environment properties. In this book chapter we categorize and examine the wide variety of ways in which these IoT sensors and actuators can support or enhance AR experiences, including quantitative insights and proof-of-concept systems that will inform the development of future solutions. We outline the challenges and opportunities associated with several important research directions which must be addressed to realize the full potential of next-generation AR.}

\section{Introduction}
\label{sec:Introduction}
While virtual content in the metaverse is designed to be immersive, it will not be experienced in isolation. In particular for augmented reality (AR), in which virtual content is integrated with our real-world surroundings, an accurate and complete understanding of the real environment is a prerequisite for high quality experiences. Obtaining this using AR devices alone is infeasible in many scenarios, raising the potential for employing external sensors placed in the surrounding environment, a form of \emph{ambient intelligence} for AR. As well as sensing the properties of an environment, it is also desirable to control them, for example to optimize the performance of core AR algorithms, or to generate stimuli in sensory modalities that are beyond the capabilities of AR devices. In this book chapter we explore the potential for wireless Internet of Things (IoT) devices to provide this type of ambient intelligence, and thereby support next-generation AR experiences in the metaverse. 

More well-studied than AR is virtual reality (VR), in which users are immersed in a fully virtual environment through visual and auditory stimuli, usually delivered via a headset (e.g., Meta Quest 2, HTC Vive), or alternatively a desktop or mobile device. To provide tactile feedback and interaction methods to the user, VR headsets are frequently used in combination with specialized controllers or other handheld devices, and less commonly with other peripherals such as gloves, body suits, helmets \cite{hoppe2021odin} and mouth accessories \cite{shen2022mouth}. There is work on integrating external devices into VR experiences to further increase the levels of immersion, beyond the capabilities of wearable devices. For example, fans have been used for representations of wind \cite{eckstein2019smart, giraldo2022towards}, while multiple works (e.g., \cite{saeidi2021exploring}) have used `climate chambers' with HVAC systems to study thermal perception in immersive virtual environments. External devices have also been employed to capture contextual data from users or the ambient environment to enhance VR experiences (e.g., \cite{rabbi2018virtual, lin2021resource}).

Here we focus on the less-studied topic of \emph{ambient intelligence to support or enhance AR}, in which virtual content is overlaid onto and integrated with a user's real-world surroundings. Specifically, we examine this in the context of mobile AR, when virtual content is presented through handheld devices such as a smartphone or tablet, or wearable devices such as a headset. As in VR, current AR devices deliver virtual content primarily in the form of visual and auditory stimuli, with limited tactile feedback sometimes available in the case of handheld devices or headsets with handheld controllers (e.g., the \mbox{Magic Leap 2}). This again raises the possibility of using external devices to adjust the real environment in other sensory modalities or to extend possible interaction methods. However, there is also an important distinction between two methods of delivering visual content in AR; handheld devices and some headsets (e.g., the \mbox{Varjo XR-3}) use video see-through (VST, sometimes referred to as video pass-through) displays, while other headsets (e.g., the \mbox{Microsoft HoloLens 2} and the \mbox{Magic Leap 2}) use optical see-through (OST) HMDs. These different designs have important implications for how the properties of real environments affect a user's perception of virtual visual content. In general, a knowledge of how the real environment affects both system functionality and human perception, along with the ability to sense and control environment properties, will enable the provision of optimal AR experiences in diverse scenarios.

Further motivation for pursuing ambient intelligence comes from the nature of next-generation AR. The coming years hold the promise of virtual content that is not only more closely integrated with our real surroundings, but which also adapts to the current context in which it is presented. This \emph{context-awareness} is key to realizing the full potential of AR; to deliver virtual content that provides a high-quality user experience, and enables optimal task performance, we require information on the specific circumstances in which it is presented \cite{grubert2017towards}. At a high level, we can break down this contextual information into environmental, user, social, and system awareness. Environmental awareness, including the environment understanding required for close integration of real and virtual content, is particularly challenging to obtain using the sensors onboard mobile AR devices alone because their view of an environment is \emph{spatially and temporally limited}; they typically only capture a small portion of an environment for a short period of time. Furthermore, due to restrictions on the quality of onboard sensors, and the fact that they are frequently in motion, the data they capture is often inaccurate. External devices on the other hand can help to address these deficiencies and generate more accurate, more complete environmental awareness, across a wider range of conditions.

\begin{figure} 
    \centering \includegraphics[width=0.66\textwidth]{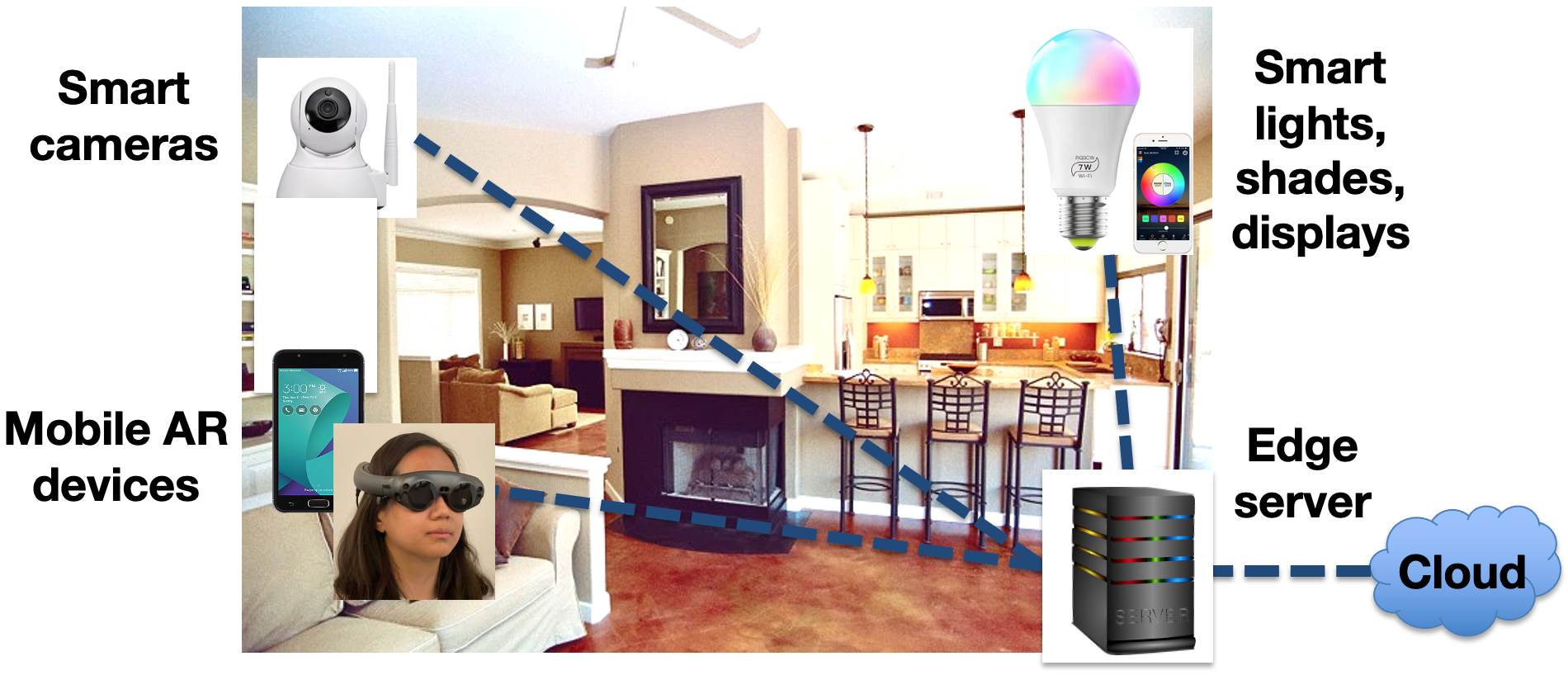}
    \caption{A high-level view of a multi-device next-generation AR architecture, in which data from mobile AR devices, wearables, and IoT sensors like smart cameras are aggregated to compute context (e.g., environment properties) on an edge or cloud server. These data are also used to inform the control of IoT actuators such as smart lights, shades, and displays, which optimize environmental conditions for AR experiences.}
    \vspace{-0.1in}
\label{fig:IoT-AR_HighLevel}
\end{figure}

The desire to both improve the accuracy and completeness of environmental awareness in AR, and control environment properties, motivates the use of Internet of Things (IoT) devices. These wirelessly connected devices have become ubiquitous across diverse settings (globally there were 11 billion active IoT devices at the end of 2021, and this is forecast to be 29 billion by 2030 \cite{morrish2022global}) and include a wide range of sensors and actuators that are suitable for detecting and adjusting environmental conditions. The widespread availability and relatively low cost of IoT sensors like smart cameras and IoT actuators like smart lights and displays, as well as the prevalence of supporting infrastructure (e.g., Wi-Fi networks), means that they can be readily leveraged for AR. Our overall vision, illustrated in Figure~\ref{fig:IoT-AR_HighLevel}, is for multi-device AR architectures, in which experiences on mobile AR devices are enhanced or supported by a set of connected devices. 
Contextual data are provided by IoT sensors such as smart cameras and wearables, while IoT actuators such as smart lights, shades, and displays optimize environmental conditions for AR, or enhance AR experiences by providing additional stimuli. An edge server or the cloud provides the storage and resources to aggregate these data, compute context, and control IoT devices.             

We define the scope of the IoT devices we cover in this book chapter through two important distinctions. Firstly, our focus is on \emph{IoT devices that support or enhance AR}, rather than simply any IoT device whose data could be displayed in AR or be controlled through an AR interface. For example, while there are interesting and useful ways in which devices such as air quality sensors and robotic arms could be combined with AR, they are excluded here. Examples of works which proposed the use of AR as a tool for visualizing IoT-generated data and interacting with IoT nodes include~\cite{jo2016ariot,michalakis2018visualizing,park2019iot,zachariah2019browsing}. Secondly, we only consider truly external or \emph{ambient} IoT devices; we exclude sensors that are attached to AR devices, such as inertial sensors or eye tracking cameras, as well as wearable sensors that may capture additional biometrics from humans in an AR environment. For a recent review of wearable sensors for AR applications, see \cite{kim2021recent}. Given the placement of ambient IoT devices in the wider environment they are particularly beneficial for environment awareness, though we note cases in which they can supply other types of context-awareness; for example, while user context data is most often captured through on-device or wearable sensors, ambient IoT cameras can also capture visual data pertinent to activity or emotion recognition \cite{scargill2022environmental}.

The contents of this book chapter are as follows. In Section~\ref{sec:RelatedWork}, we cover related work on combining AR and IoT devices, methods of communication between them, and a network architecture that will support the implementation of ambient intelligence for AR, edge computing. In Section~\ref{sec:AmbientIoTforAR}, we categorize the different ways in which ambient IoT devices can support or enhance AR experiences, including relevant sensors and actuators for each use case; then in Sections~\ref{sec:ARIoTSensing} and Sections~\ref{sec:ARIoTActuation}, we discuss in more detail the possibilities for IoT sensors and actuators respectively, organized by use case. In Section~\ref{sec:ChallengesandResearchDirections} we cover open challenges and research directions, and in Section~\ref{sec:Conclusion} we provide a conclusion.

\section{Related Work} 
\label{sec:RelatedWork}
In this section we review existing work on systems which incorporate both AR and IoT devices, different methods of communication that have been used to connect AR and the IoT, and a network architecture that will support the implementation of ambient intelligence for next-generation AR. \hfill \break   

\textbf{Systems incorporating AR and IoT devices: }AR can be used to enhance interactions with IoT devices, including both the visualization of IoT data and the provision of an immersive interface to access and control IoT devices. For example, in \cite{phupattanasilp2019augmented} the authors described an AR-IoT system that displays real-time IoT data as holographic content to enhance object interactions; this system is applied to crop monitoring, to provide object coordinates of plants and information collected from IoT devices such as the fertilizers used. Similarly, Sun et al. \cite{sun2019magichand} presented MagicHand, an AR system that allows users to interact with IoT devices by detecting, localizing, and enabling augmented hand controls; the system was implemented using a 2D convolutional neural network (CNN) and achieved high gesture recognition accuracy.
In \cite{sun2019towards} IoT sensor data were overlaid onto industrial machines using AR, with more accurate pose estimates (and thereby better aligned overlays) obtained by applying deep learning to RGB and depth images of the machine. Visualizing and identifying IoT objects using an AR interface has also been shown to improve shopping experiences, by increasing perceived usability and satisfaction in user interactions \cite{jo2019iot+}. \hfill \break

\textbf{Communication between AR and IoT devices: }Prior works have demonstrated how AR devices can communicate with IoT devices through the Internet; examples include an AR application that displays agricultural data from temperature or moisture sensors for crop monitoring \cite{macias2011browsing}, and a web-based AR application framework to visualize the state changes of a coffeemaker through a visual tag \cite{leppanen2014augmented}. Others have highlighted the importance of scalable systems with efficient data management for AR applications \cite{chatzopoulos2017mobile, qiao2019web}; to this end the authors of \cite{zachariah2019browsing} proposed an AR-based browsing architecture that identifies new IoT devices and enables immediate interactions with easily controllable user interfaces. Similarly, Blanco-Novoa et al. proposed a framework that allows AR and IoT devices to communicate dynamically in real time through standard and open-source protocols such as MQTT, HTTPS, or Node-RED \cite{blanco2020creating}. Developing systems which incorporate AR devices into a network of IoT devices remains an important topic of research; for example, VisIoT \cite{park2019iot} supports tracking and visualizing the location of IoT nodes in real time through a combination of data collected from camera, inertial measurement unit (IMU), and radio frequency (RF) modules, while Scenariot \cite{huo2018scenariot} integrates the discovery and localization of IoT devices into an AR system by embedding RF modules into both AR and IoT devices. These works highlight the need for further research on device localization and calibration, as well as system scalability (with respect to e.g., bandwidth consumption), to inform the development of IoT-supported AR systems.\hfill \break

\textbf{Edge computing for AR: }In order to leverage the large amounts of data from AR and ambient IoT devices, context-aware AR systems will require storage and computational resources beyond the constraints of these devices alone. Given the low latency requirements of many aspects of context-aware AR, along with the privacy concerns associated with transferring sensitive information about users or environments to the cloud, many have deemed edge computing a particularly promising network architecture~\cite{liu2020collabar,Yongjie2022deep,Zhao2021Xihe,zhang2022sear,wang2022leaf+}. In this architecture, a server is placed physically close to mobile AR devices and ambient IoT devices (e.g., in the same building), helping to address the aforementioned latency and privacy requirements. Existing work has already demonstrated the benefits of offloading tasks to the edge for various aspects of AR system functionality, including elements of SLAM pipelines \cite{ali2020edge, wen2020edge, xu2020edge}, lighting estimation \cite{Zhao2021Xihe} and object detection \cite{liu2019edge, ran2018deepdecision, herrmann2021object}, and in our ongoing work we have developed multiple edge architectures for context-aware AR. For example, in \cite{scargill2021will} we developed a system to predict the quality of virtual content positioning (a function of AR device pose estimate error) from environment properties, in which we transmitted data collected on the AR device to the edge for the computationally expensive pre-processing and model inference. In \cite{liu2020collabar} we presented an edge-assisted collaborative image recognition system, in \cite{scargill2022catch} we demonstrated an edge-supported AR application that analyzed user eye movements to recognize common activities in a museum scenario, and we have developed multiple systems that provide edge-based provisioning of contextual virtual content \cite{glushakov2020edge}. We see the incorporation of IoT devices that provide additional contextual data as a natural extension to these edge architectures for context-aware AR, and we recently presented an example of this in \cite{scargill2022iot} (see Section~\ref{subsec:ActuationSpatialUnderstanding} for further details).

\section{Ambient IoT for AR}
\label{sec:AmbientIoTforAR}
In this section, we categorize the different ways in which ambient IoT devices can support or enhance AR experiences (Section~\ref{subsec:UsesofAmbientIoTforAR}). We then provide an overview of the different types of IoT devices that may be employed, along with their associated uses, and examples of their use for AR or VR (Section~\ref{subsec:AmbientIoTDevicesforAR}).

\subsection{Uses of Ambient IoT for AR}
\label{subsec:UsesofAmbientIoTforAR}
Central to next-generation AR is the incorporation of context-awareness -- adapting virtual content according to the environmental, user, social, and system context in which it is presented. Ambient IoT sensors are able to collect environment data that is more accurate and complete than AR devices alone, while ambient IoT actuators can be used to adjust environment properties for better system performance or a higher quality user experience. In order to categorize the ways in which detecting and adjusting environment properties using IoT devices can support or enhance AR, here we define five high-level aspects of AR which contribute to the quality of a user's experience:

\begin{myitemize}
  \item \emph{Spatial understanding} concerns information about the physical properties of the environment, which includes representations of the real-world surfaces present, the detection of fiducial or image-based markers, and real-time pose estimates for the AR device; this information is required for accurate spatial registration of virtual content. 
  \item \emph{Semantic understanding} takes this environmental awareness a step further, and provides a knowledge of the type, poses, and states of objects and surfaces present, which may be used to inform spatial understanding or display context-aware content.  
  \item \emph{Contextualized content} refers to the ways in which either the spatial or semantic understanding obtained through IoT devices, or environment properties such as light and visual texture directly, may be used to inform adaptations to virtual content.
  \item \emph{Interaction} covers how current interaction methods in AR (e.g., hand gestures, eye tracking) may be enhanced or extended using IoT devices.
  \item \emph{Immersion} relates to how IoT actuators can be used to increase an AR user's sense of immersion (i.e., the sense that virtual content is truly present in their real environment).
\end{myitemize}

\subsection{Ambient IoT Devices for AR}
\label{subsec:AmbientIoTDevicesforAR}
In Table~\ref{tab:IoTforAROverview}, we present an overview of ambient IoT devices that may be used to support or enhance AR experiences. We group included devices by the type of information they collect from or convey to AR devices or users (e.g., visual, auditory), rather than the underlying functionality of the device. For example, many motion sensors detect changes in thermal energy, but given that in this case they provide information regarding the state of the visual environment (i.e., human presence), we classify them here under `visual'. Similarly, strain gauges, which detect changes in electrical resistance, are classified as visual because they provide information about the deformation of an object, effectively enhancing the visual perception and acuity of an AR user. For each type of information, we divide IoT devices into sensors or actuators, list example devices, state their possible uses from the categories we defined in Section~\ref{subsec:UsesofAmbientIoTforAR}, and provide examples of related work for these use cases. 

\begin{table}
  \centering
  \caption{Overview of ambient IoT devices (sensors and actuators) with potential uses for AR. Sensors and actuators are grouped by the type of information they collect from or convey to AR devices or users (e.g., visual, auditory).}
  \begin{tabular}
  {|>{\centering\arraybackslash}m{1.8cm}||
  >{\centering\arraybackslash}m{2.2cm}|
  >{\centering\arraybackslash}m{3.0cm}|
  >{\centering\arraybackslash}m{2.1cm}||
  >{\centering\arraybackslash}m{2.2cm}|
  >{\centering\arraybackslash}m{3.0cm}|
  >{\centering\arraybackslash}m{2.1cm}|
  }
  \hline
\multirow{3}{1.8cm}{\centering \textbf{Type of information provided}} & \multicolumn{3}{c||}{\textbf{Sensors}} & %
    \multicolumn{3}{c|}{\textbf{Actuators}} \\
\cline{2-7}
 & \textbf{Devices} & \textbf{Uses} & \textbf{Examples of use for AR or VR} & \textbf{Devices} & \textbf{Uses} & \textbf{Examples of use for AR or VR} \\
\hline
Visual&Cameras, depth sensors, motion sensors, light sensors, strain gauges, pressure sensors&Spatial understanding, semantic understanding, contextualized content, interaction&\cite{minh2019motion, masai2020face, yang2020ear, liu2018gesture, rohmer2014interactive, scargill2021will}&Light bulbs, projectors, electronic displays (e.g., LCD, E-Ink)&Spatial understanding, semantic understanding, interaction, immersion&\cite{scargill2022iot, ahuja2019lightanchors}\\
\hline
Auditory&Microphones&Semantic understanding, contextualized content, interaction&\cite{ghorbani2023decision, morotti2020fostering}
&Speakers&Immersion&\cite{lee2017location}\\
\hline
Haptic&Wind sensors, physical buttons and proxies, touchscreens&Contextualized content, interaction&\cite{kim2018blowing, baumeister2017cognitive, englmeier2020tangible}
&Haptic surfaces, fans&Interaction, immersion&\cite{eckstein2019smart, giraldo2022towards}\\
\hline
Olfactory&Nanomechnical sensors&Semantic understanding&-&Diffusers, olfactometers&Immersion&\cite{archer2022odour}\\
\hline
Thermal&Thermocouples, resistance temperature detectors, infrared temperature sensors&Semantic understanding, contextualized content&\cite{bonanni2005attention, natephra2019live, phupattanasilp2019augmented}
&Heaters, HVAC systems&Immersion&\cite{eckstein2019smart, saeidi2021exploring}\\
\hline
\end{tabular}  \label{tab:IoTforAROverview}
\end{table}

Considering first IoT sensors, there is a wide variety of useful information we can collect from the surrounding environment, and we identify four key ways in which these data may be leveraged. Firstly, we can enhance the spatial understanding, semantic understanding, and interaction capabilities of AR systems by sensing from additional and more advantageous poses. Secondly, with sufficient data on how environmental conditions impact AR systems, we can predict the current level of performance and overall quality of an AR experience. Thirdly, we can use the data collected from sensors to adapt AR system functionality or the presentation of virtual content for the current environment. Finally, we can contextualize virtual content, in that it is related to or reacts to current conditions. Throughout Section~\ref{sec:ARIoTSensing}, we discuss in more detail how IoT sensors can be used to support or enhance AR.  

Beyond detecting the properties of the real environment using IoT sensors, we can also \emph{adjust} the properties of the real environment using IoT actuators. The motivation for this is threefold. Firstly, because key aspects of AR system performance and user perception are affected by environment properties, we can optimize environments to achieve the best possible performance. For example, the accuracy of spatial understanding is dependent on sufficient levels of light and visual texture, light levels affect the performance of algorithms related to semantic understanding and interaction, and human perception of virtual content is also affected by the properties of both light and textures. Secondly, we can improve the quality of a user's interactions in AR, by providing alternative interaction methods (e.g., electronic displays) or enhancing existing ones (e.g., adding tactile feedback using haptic surfaces). Thirdly, we can enhance a user's sense of immersion in an AR experience. This may involve the generation of visual and auditory stimuli to extend what is possible on AR devices (using e.g., light projectors or speakers), or the generation of sensory stimuli in modalities that cannot be generated on AR devices (using e.g., fans, diffusers, or heaters). We discuss the possibilities for ambient IoT actuators in more detail in Section~\ref{sec:ARIoTActuation}.

\section{IoT-based Sensing for AR}
\label{sec:ARIoTSensing}
In this section we discuss in more detail how ambient IoT sensors could be used to support or enhance AR experiences. Each subsection examines a different use which we defined in Section~\ref{subsec:UsesofAmbientIoTforAR} and listed in Table~\ref{tab:IoTforAROverview}; for sensors we cover spatial understanding (Section~\ref{subsec:SensingSpatialUnderstanding}), semantic understanding (Section~\ref{subsec:SensingSemanticUnderstanding}), contextualized content (Section~\ref{subsec:SensingContextualizedContent}), and interaction (Section~\ref{subsec:SensingInteraction}).   
\subsection{Spatial Understanding}
\label{subsec:SensingSpatialUnderstanding}
A fundamental component of all AR systems which position virtual content relative to the real world is spatial understanding. Indeed an accurate and detailed knowledge of our physical surroundings is essential for next-generation AR systems which aim to closely integrate the real and virtual worlds. In this subsection, we first provide background information on the techniques behind spatial understanding in AR. We then examine how environment properties affect spatial understanding, and hence why obtaining knowledge of these properties through IoT sensors is useful, before exploring how IoT sensors may be used directly in spatial understanding algorithms.

\subsubsection{Background on Spatial Understanding in AR}
\label{subsec:BackgroundSpatialUnderstanding}

Understanding one's physical surroundings is a fundamental component of both marker-based and markerless AR, in order to accurately overlay virtual content onto a view of the real-world environment. A marker-based AR system uses a marker, commonly a paper-printed static image with distinct features, to obtain information about the position and orientation of an object in the surrounding space. On the other hand, a markerless AR system is based on the use of simultaneous localization and mapping (SLAM) to understand the surrounding space without the use of markers.\hfill \break

\textbf{Marker-based AR: }Marker detection based on image processing has been a popular approach to enable marker-based applications in AR due to its ease of use and accurate tracking of an object \cite{garrido2014automatic}. A multitude of markers with different patterns and features has been used in marker-based AR systems, including binary fiducial markers (e.g., ARToolkit \cite{ARtoolkit}, ArUco \cite{Aruco}, ARTag \cite{fiala2009designing}), a variation of fiducial markers specialized for robotics applications (e.g., AprilTag \cite{olson2011apriltag}), and image-based markers with a high number of unique feature points (e.g., Vuforia \cite{Vuforia}). The detection of these markers is based on the processing of the image frames captured by the camera on an AR device -- the pose of the device is estimated by finding contours, features, or lines \cite{hirzer2008marker} within the marker. However, the accuracy and reliability of marker detection in AR are largely determined by the performance of the camera capturing images of the scene and the environmental conditions of the scene where the marker is located. Poor camera calibration or focus, or low image resolution, can potentially result in low  pose estimation accuracy of the marker in the scene \cite{zhang2002visual}. Additionally, environmental properties such as lighting or the distance from the camera to the marker are other factors that can affect pose estimation accuracy. We discuss the use of IoT sensors and actuators to address challenges related to marker detection in Section~\ref{subsubsec:ObjectStateIoTSensors} for the use of IoT sensors in object state detection, and in  Section~\ref{subsubsec:EnvironmentOptimizationMarkerBased} for environment optimization using IoT actuators. \hfill \break

\textbf{Markerless AR: }
SLAM is a key enabling technology for markerless AR. Visual and visual-inertial SLAM (V-SLAM and VI-SLAM), using cameras either alone or in combination with inertial sensors, have demonstrated remarkable progress over the last three decades~\cite{cadena2016past}.
Due to the affordability of cameras and the richness of information provided by them, V-SLAM using monocular~\cite{rublee2011orb,mur2017orb}, RGB-D~\cite{mur2017orb}, and stereo~\cite{mur2017orb} cameras has been widely studied. 
To provide robustness to textureless areas,  motion blur, illumination changes, there is a growing trend of employing \mbox{VI-SLAM}, that assists cameras with an IMU~\cite{weiss2012real,qin2018vins,campos2021orb}; VI-SLAM has become the de-facto standard SLAM method for 
modern augmented reality platforms~\cite{ARCore,ARKit}. In VI-SLAM,  visual information is fused with IMU data to achieve more accurate and robust localization and mapping performance~\cite{weiss2012real,qin2018vins,campos2021orb}. Due to the high computational demands incurred by V- and VI-SLAM on mobile devices, offloading parts of the workload to edge servers has recently emerged as a promising solution for lessening the loads on mobile devices and improving overall performance~\cite{xu2020edge, ali2020edge,  xu2022swarmmap,Chen2023AdaptSLAM}.  A standard
approach~\cite{ali2020edge,  xu2022swarmmap,Chen2023AdaptSLAM} is to offload computationally expensive tasks, such as 
place recognition, the process of taking
a sensor snapshot (e.g., an image) of a location and  querying it in a large, geotagged database gathered from prior measurements,
and loop closings, the process of determining whether AR users are revisiting the same place.
Both the aforementioned papers~\cite{qin2018vins,campos2021orb,xu2020edge, ali2020edge, xu2022swarmmap,Chen2023AdaptSLAM} and commercial AR devices that employ markerless AR (e.g., Android devices with ARCore \cite{ARCore}, iOS devices with ARKit \cite{ARKit}, or headsets such as the Microsoft HoloLens 2 \cite{HoloLens}) implement VI-SLAM using sensor data captured onboard mobile devices, and these data are both spatially and temporally limited. To address this limitation, in Section~\ref{sec:EnhancingSpatialUnderstanding} we discuss methods to increase the accuracy and completeness of spatial understanding by integrating the sensor data obtained onboard AR devices with data obtained from IoT sensors.

\subsubsection{Estimating the Quality of Spatial Understanding Using IoT Sensors}
\label{subsubsec:EstimatingtheQualityofSpatialUnderstandingUsingIoTSensors}
Because of the role of vision-based sensing in spatial understanding in AR, and the nature of VI-SLAM algorithms in particular, the properties of the visual environment impact the accuracy and completeness of spatial understanding. Therefore a thorough knowledge of those properties, obtained through ambient IoT sensors, is highly useful in identifying problematic regions for tracking, estimating current levels of spatial understanding, and informing system adaptations for current environmental conditions. Until recently, knowledge of the impact of environmental properties was limited to qualitative guidelines; however, our recent work \cite{scargill2022integrated, scargill2022environmental} provides quantitative insights on the impact of both light and visual texture on VI-SLAM performance.

To obtain these quantitative insights on the impact of environmental conditions on VI-SLAM performance, we developed and implemented two methodologies. In one \cite{scargill2022integrated}, we measure the pose estimate error of open-source VI-SLAM algorithms (e.g., ORB-SLAM3 \cite{campos2021orb}) using a game engine-based emulator; we use the trajectories from existing VI-SLAM datasets (e.g. TUM VI~\cite{schubert2018tum}, SenseTime \cite{jinyu2019survey}) to create new camera images in realistic virtual environments, and combine that with the original inertial data from those datasets. In the other \cite{scargill2021here}, we measure virtual object position error (determined by pose estimate error) on commercial AR platforms (e.g., ARCore \cite{ARCore}, ARKit \cite{ARKit}), by aligning virtual objects with a real world reference point using our open-source AR app. Our game engine-based methodology facilitates fine-grained control of environment properties (i.e., the exact properties of the light sources and textures in a virtual environment), while our methodology for commercial AR supports monitoring of environment conditions with either AR device sensors or ambient IoT sensors.\hfill \break

\begin{figure*}[htp]
\centering
   \subfloat[SenseTime A1]{\label{fig:ConcreteLightA1}\includegraphics[width=.48\textwidth]{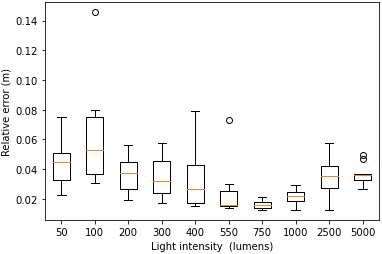}}
   \hspace{\fill}
   \subfloat[SenseTime A4]{\label{fig:ConcreteLightA4}\includegraphics[width=.48\textwidth]{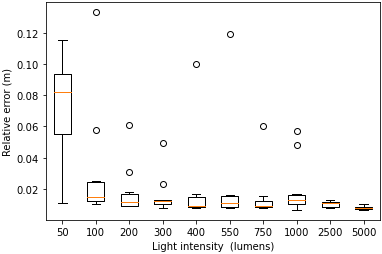}}
   \caption{VI-SLAM pose estimation error for two SenseTime \cite{jinyu2019survey} trajectories in a 6m$\times$6m$\times$4m virtual concrete room \cite{scargill2022integrated}, showing the translational component of relative error when different light intensities are emitted by a light source (100 trials, 10 at each light intensity). Optimal light intensity for VI-SLAM performance depends on the structure, textures, and reflectance properties of an environment, as well as the camera trajectory. While a medium light intensity will be optimal when specular reflections due to high illuminance are a factor (e.g., Figure~\ref{fig:ConcreteLightA1}), trajectories with views of environment regions where light sources are distant or occluded will result in optimal performance at high light intensities (e.g., Figure~\ref{fig:ConcreteLightA4}). }\label{fig:ConcreteLight}
\end{figure*}

\textbf{Light: }Illuminance, the amount of light incident on environment surfaces per unit area, determines the accuracy with which environment surfaces can be mapped or tracked, because it determines the extent to which visual features are detectable for tracking. We illustrate an example of this in Figure~\ref{fig:ConcreteLight}; in these experiments we used our game engine-based emulator to run two SenseTime \cite{jinyu2019survey} trajectories in a 6m$\times$6m$\times$4m virtual concrete room, with 10 different overhead light intensities \cite{scargill2022integrated}, and 10 trials per light setting. We then plotted VI-SLAM pose estimation performance (in terms of relative error, the translational component of relative pose error) against the 10 light intensity settings. Figure~\ref{fig:ConcreteLightA1} shows the results for SenseTime A1, a trajectory involving slow side-to-side motion facing a wall (as if inspecting a virtual object at head height), followed by repeated walking away and returning with the camera angled more towards the floor (described in \cite{jinyu2019survey} as `inspect+patrol'). Figure~\ref{fig:ConcreteLightA4} shows the results for SenseTime A4, with slow motion focused on a small area of the floor, followed by the same slow side-to-side motion facing the wall (described as `aiming+inspect'). 

For SenseTime A1, optimal performance is obtained at a medium light intensity (750 lumens), at which there is sufficient light to ensure recognizable features are visible, but those features are not obscured by specular reflections -- this is particularly a factor during the `walking away and returning' portion of the sequence. For the less challenging inertial data in SenseTime A4, performance is largely determined by the illuminance of the small area of floor at the start of the sequence, and specular reflections are not a major factor; as such performance is poor at low light levels when the small area of floor is too dark, and optimal performance is obtained at the highest light intensity. These results illustrate that optimal environment illuminance differs depending on which regions of an environment user trajectories cover, and that monitoring of illuminance in specific environment regions of interest will be informative for the estimation of current VI-SLAM performance. 

We envision multiple IoT ambient light sensors being employed to accurately measure lighting properties in different parts of the environment, without requiring an AR device. This will enable both the proactive identification of environment regions where lighting may cause tracking errors, and a more complete understanding of how lighting conditions change over time. Data on the position of virtual content or user trajectories will inform the most relevant positions for ambient light sensors to be placed. The output from these sensors is also highly useful for environment optimization systems which control properties such as illuminance based on occupant needs, as we show in Section~\ref{subsec:ActuationSpatialUnderstanding}.\hfill \break

\begin{figure} 
    \centering \includegraphics[width=0.80\textwidth]{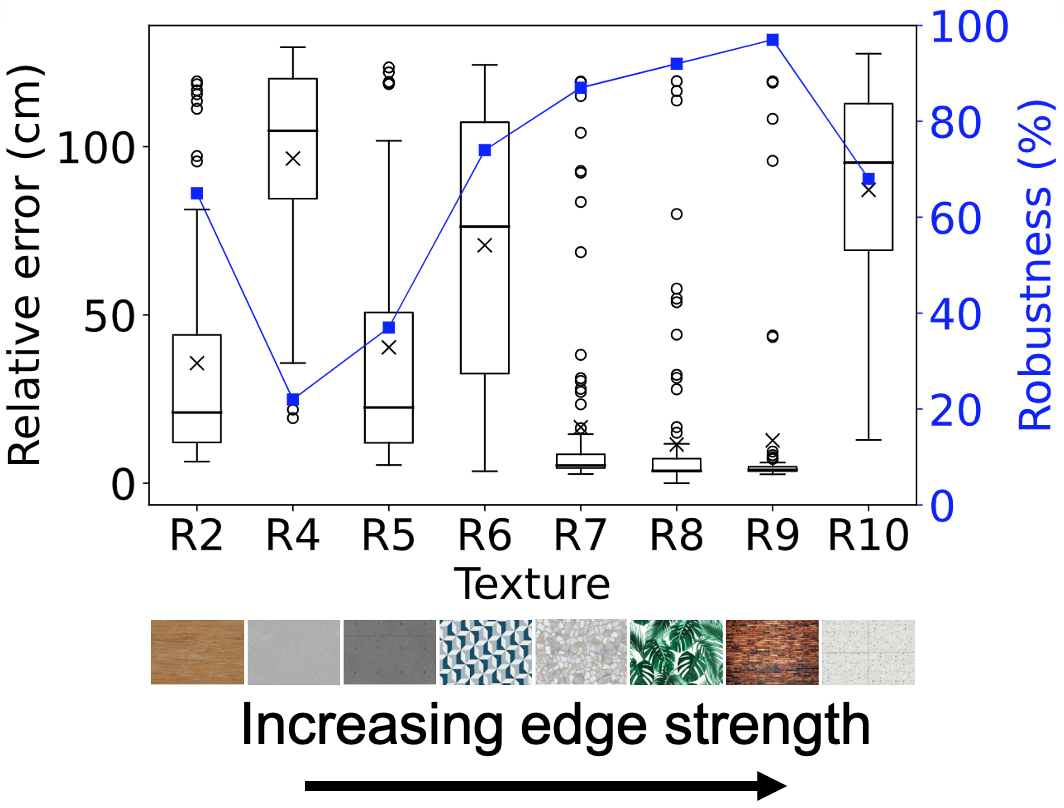}
    \caption{VI-SLAM pose estimate performance for TUM VI room5 \cite{schubert2018tum} with various visual textures, ordered by the edge strength of the texture. The interior of a 6m$\times$6m$\times$4m virtual room was covered in each texture, and 100 trials were conducted for each (10 at each of 10 different light levels, light levels shown in Figure~\ref{fig:ConcreteLight}). Performance was measured using relative error, the translational component of relative pose error, and robustness, the mean percentage of tracked frames over all trials.}
    \vspace{-0.1in}
\label{fig:VisualTexture_Real_room5}
\end{figure}

\textbf{Visual texture: }Given the reliance of feature-based SLAM on recognizable visual textures in the surrounding environment, estimating the properties of visual textures present is also a valuable predictor of device tracking performance in markerless AR. In quantitative experiments on a state-of-the-art open-source VI-SLAM algorithm (ORB-SLAM3 \cite{campos2021orb}) using our game engine-based emulator, we showed that both the edge strength and complexity of a texture, as well as how a texture is impacted by motion blur, affect pose estimate error magnitude \cite{scargill2022integrated}. For example, Figure~\ref{fig:VisualTexture_Real_room5} shows the VI-SLAM pose estimation performance (in terms of relative error, the translational component of relative pose error, and robustness, the mean percentage of tracked frames) we obtained when running the `room5' trajectory from the TUM VI dataset \cite{schubert2018tum} in 6m$\times$6m$\times$4m cuboid environments covered with different visual textures. The textures are ordered according to their edge strength (variance of the Laplacian), with higher numbers indicating higher edge strength values. Three out of four of the textures with high edge strength, stone (R7), plants wallpaper (R8), and brick (R9) resulted in median relative error $\leq$5cm, compared to $>$20cm for all other textures. The notable exception was the speckled marble texture (R10), with a median relative error of 95cm. In this case, the fine texture was greatly affected by motion blur, resulting in less recognizable texture (low edge strength) in camera images from dynamic portions of the trajectory.

\begin{figure} 
    \centering \includegraphics[width=0.66\textwidth]{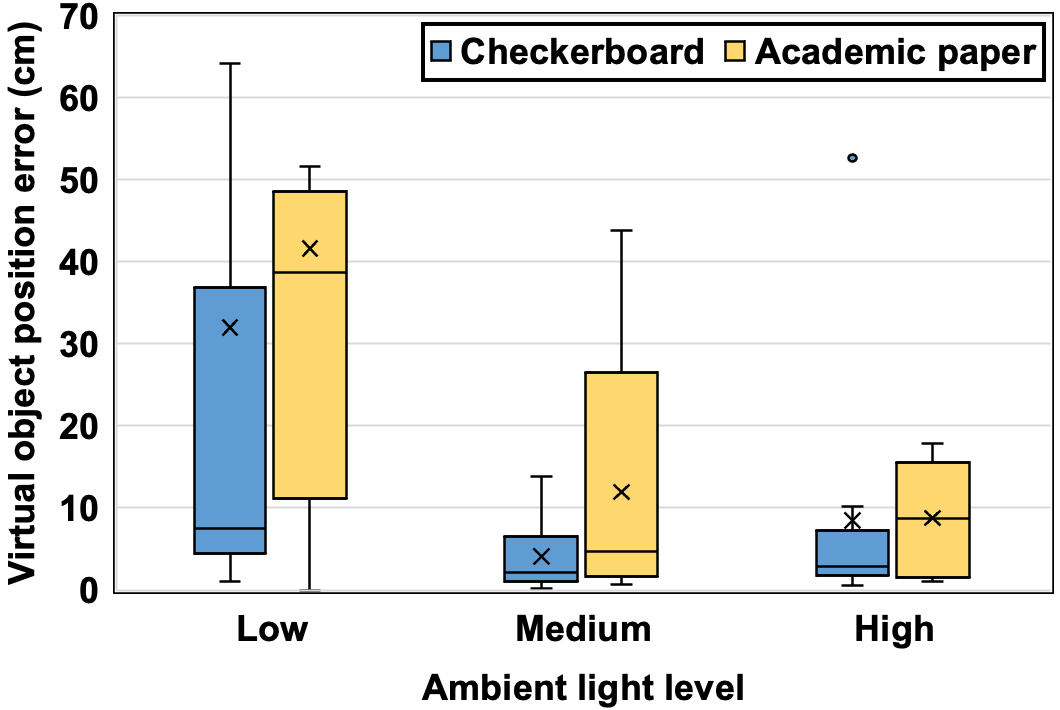}
    \caption{Virtual object position error on the Samsung Galaxy Note 10+ smartphone after walking away approximately 7m and returning, when the virtual object was placed on a checkerboard or academic paper texture, with a low (50--100 lux), medium (150--450 lux) or high (500--1000 lux) ambient light level. The fine texture of the academic paper is less robust to lower light levels due to it being more easily obscured by noise in the AR device camera image \cite{scargill2022iot}.}
    \vspace{-0.1in}
\label{fig:IoT_Drift}
\end{figure}

In our experiments on commercial AR platforms we have demonstrated that in terms of virtual object position error (a function of pose estimate error magnitude), some visual textures are more robust to low illuminance than others \cite{scargill2022iot}. In these experiments we measured virtual object position error, the 3D Euclidean distance between where a virtual object was originally placed and where it appears after walking away approximately 7m and returning, in different environmental conditions using our open-source app \cite{scargill2021here}. We conducted our experiments in a university lab, and tested two textures where the virtual object was placed, a checkerboard and an academic paper, at three ambient light levels (low, 50-100 lux; medium, 150-450 lux; high, 500-1000 lux), with 10 trials for each of the six settings. Figure~\ref{fig:IoT_Drift} shows our results for these experiments on the Samsung Galaxy Note 10+ smartphone (ARCore v1.28). Our results illustrate how error increases at lower ambient light levels, but that the checkerboard texture was more robust to this effect than the academic paper (mean errors of 4.1cm and 12.0cm respectively at the medium light level). We observed that at the medium light level, noise in smartphone camera images has minimal effect on the checkerboard texture, but obscures the finer texture of the academic paper, making VI-SLAM-based place recognition more challenging, and resulting in greater error.

These relationships between tracking quality and visual texture motivate the proactive monitoring of texture properties; images of the environment periodically captured by IoT cameras can be transmitted to an edge server for the required image processing. Again, not only does this allow us to identify and manually address problematic regions that might cause poor tracking quality before they are encountered by AR devices, but we can automatically optimize illuminance or texture based on the current types of visual texture present (see Section~\ref{subsubsec:EnvironmentOptimizationforMarkerlessAR}). Another promising direction is implementing a form of adaptive SLAM by adjusting feature extraction parameters based on visual texture; for example, it may be beneficial to lower corner detection thresholds when low-contrast textures are detected.

\subsubsection{Enhancing Spatial Understanding Using IoT Sensors}
\label{sec:EnhancingSpatialUnderstanding}
As well as measuring environment conditions to estimate current levels of spatial understanding, we can also use the data from ambient IoT sensors as input to spatial understanding algorithms directly. There are several promising possibilities in this area, which we discuss below.\hfill \break    

\textbf{Collaborative SLAM: }The visual data obtained from cameras onboard AR devices is both spatially and temporally limited, and frequently suffers from distortions such as noise or motion blur. To help alleviate some of these issues, we can additionally employ the visual data from ambient IoT cameras. This use of multiple vantage points can be seen as a type of \emph{collaborative SLAM}, where captures of different devices can be combined to obtain a higher-quality overall map, that leads to higher-quality pose estimation by the AR device~\cite{karrer2018cvi,schmuck2019ccm,xu2022swarmmap}. To further 
extend the sensing capabilities
using external sensors, in our planned work we will use IoT sensors (e.g., a surveillance camera) located in the vicinity of AR devices to provide extra information for pose estimation in SLAM. Specifically, we will use deep learning-based object detection
and person re-identification for obtaining semantic information, that is, localizing and tracking users wearing mobile AR devices in the surveillance camera frames. We will fuse the semantic information obtained from the surveillance camera with VI-SLAM running onboard AR devices to
achieve more robust and accurate pose estimation.\hfill \break

\textbf{Collaborative depth mapping}: In addition to VI-SLAM, modern AR devices are increasingly relying on time-of-flight \emph{depth sensors} to aid with spatial understanding, and sensing from multiple vantage points can also be advantageous for obtaining accurate depth maps. Time-of-flight sensors struggle to obtain reliable data when the observed scene contains materials with low reflectivity, strongly specular objects, and reflections from multiple objects~\cite{tofdesignguide,tofprinciples}. Not only that, but indirect time-of-flight depth sensors such as those on the Microsoft HoloLens 2, Magic Leap 2 and some high-end Android smartphones are limited in range, while the direct time-of-flight depth sensors (marketed as LiDAR) on high-end iOS devices produce sparse readings that may be inaccurate at shorter distances. This leads to depth maps that are missing valid estimates in large parts of a frame, and incomplete spatial understanding.   

In our evaluation of depth estimates obtained by a Microsoft HoloLens~2 (in the long throw mode) across a range of representative indoor environments, we found that on average 30\% of depth pixels in a frame were missing~\cite{zhang2022indepth}. We also collected an indoor dataset of 18.6K depth maps on a Samsung Galaxy Note 10+ smartphone, of which 58\% had greater than 40\% missing pixels~\cite{zhang2022indepth}. Based on the properties of these missing pixels, we determined that the range of indirect time-of-flight depth sensors on current AR devices is approximately 5m. We also observed that smaller angles ($60^{\circ}$ or less) between the sensor's optical axis and the target surface resulted in large numbers of missing pixels. These problematic conditions are prevalent in AR scenarios -- in larger rooms a distance of greater than 5m between an AR device and the nearest surface is common, while a depth sensor naturally faces outward toward a wall (due to how a smartphone is usually held or a headset is worn), but target surfaces in AR are often horizontal planes such as a table.

Ambient IoT depth sensors can help to increase the completeness of raw depth maps by capturing data from poses that are not accessible to or normally covered by AR devices. To address limitations in sensor range, ambient depth sensors can be positioned closer to surfaces that are only observed from large distances by AR devices. Horizontal planes, frequently incomplete due to their similar angle of orientation to the optical axis of AR device sensors, can be captured by downward-facing IoT depth sensors. Even some challenging reflections may be avoided from different viewpoints. We envision this collaborative sensing approach being combined with existing techniques for depth map completion such as~\cite{zhang2022indepth,zhang2018deepdepth,merrill2021icra,deeplidar}, with the more complete depth data from multiple sensors combined on the edge server for a less challenging depth inpainting task.\hfill \break 



\textbf{Scene change detection: }Ambient IoT cameras can also be used to establish whether the environment has changed between different AR sessions, to trigger SLAM remapping as required to improve the quality of spatial scene understanding (and conversely to avoid unnecessary time- and resource-consuming remapping if the environment did not change). Scene change detection based on stationary cameras' inputs is a long-examined, well-formulated problem, for which many solutions have been proposed~\cite{wang2014cdnet}. Extending existing solutions to incorporate the specific constraints of heterogeneous multi-device platforms we envision (IoT and AR, stationary and mobile, devices), for the specific case of scene change detection in the context of VI-SLAM, has the potential to significantly reduce the extent of mapping that would be required to achieve high-quality, spatially-aware AR experiences.\hfill \break

\subsection{Semantic Understanding}
\label{subsec:SensingSemanticUnderstanding}
A key element of next-generation AR is the use of semantic algorithms that detect the type, position, orientation, and even the current state of objects and surfaces within an environment. Not only can they be combined with SLAM to extend or enhance spatial understanding (e.g.,~\cite{zhang2020flowfusion}), but they also enable the delivery of various types of content to the user. Firstly, directly annotating a visual display with semantic information has a wide variety of applications, from language learning \cite{huynh2019situ} to firefighting \cite{bhattarai2020embedded}. Secondly, this knowledge can be used to inform user interactions, e.g., suggesting appropriate places for the user to position virtual content, or objects that can be interacted with in a specific application. Finally, semantic understanding also enables the provision of more intelligent content, such as avatars or virtual characters that interact naturally and autonomously with real-world objects. Here we focus on the topic of object detection to illustrate the role of ambient IoT sensors, however the techniques we describe may also be applied to other types of semantic understanding, such as semantic segmentation.\hfill \break

\subsubsection{Background on Object Detection in AR}
\label{subsubsec:BackgroundonObjectDetectioninAR}
By running object detection models on images captured by AR devices, we can detect the type, pose and extents of common objects that are present in real world environments, which provides us with a more in-depth understanding of environmental context and informs the rendering of virtual content.
Although current advancements in deep neural networks (DNNs)
have shown superior performance in object detection~\cite{apicharttrisorn2019frugal, li2020object, hu2020object, le2021augmented},
executing large networks on computation-constrained devices such as AR devices and IoT sensors with low latency remains a challenge.  
To address this, edge-supported architectures are needed to offload computation from the AR devices and IoT sensors and improve the end-to-end latency~\cite{zhang2017networking,wang2022leaf+,liu2020collabar,Yongjie2022deep,xu2022tutti}. 
As the pervasive deployment of mobile AR will offer numerous opportunities for multi-user collaboration, prior works have also studied object detection that exploits the visual information captured by different AR devices~\cite{choudhary2020multi,zhang2022sear}. Nevertheless, collaborative object detection for AR devices and IoT sensors, where visual information is captured from highly distinct vantage points, is an unexplored area of research. The depth information that is obtained by specialized AR headsets and high-end smartphones equipped with time-of-flight depth sensors may also be employed to aid in object detection \cite{ophoff2019exploring}, while recent work has investigated the use of point clouds generated on these types of AR devices as input to object detection models \cite{choudhary2020multi,herrmann2021object}. 

Another approach is to detect and track objects in the environment by matching current input data -- either 2D feature points in captured images, or the 3D data in a generated point cloud -- with a predefined reference. In the case of images, the marker detection techniques we described in Section~\ref{subsec:BackgroundSpatialUnderstanding} may be used by attaching a printed marker to a real-world object. Alternatively, feature points can be matched across input and reference images \cite{eom2022armagnifier, lee2017object}. By comparing the features extracted on reference images (i.e., images of objects that need to be detected or tracked) to the features detected on input images (i.e., images of AR scenes where the desired objects are present) using descriptors such as ORB, SIFT, or SURF \cite{marchand2015pose}, the pose estimation of a desired object can be found by computing the homography matrix. Efforts to enhance this markerless registration through improved feature matching include \cite{chen2016improved, lee2008hybrid, kurz2011gravity}.
Similarly for point clouds, input data is matched with the 3D points captured in a prior scan of the object, through point cloud registration \cite{huang2021comprehensive}. This approach is used to enable the detection of previously scanned objects on ARKit \cite{ARKitObjectDetection}. In general, feature matching approaches require fewer computational resources when compared to neural-network-based object detection, but are less robust to environmental factors such as lighting, image distortion, the distance of the AR device to an object, and background textures.

\subsubsection{Enhancing Object Detection Using IoT Sensors}
\label{subsec:SensingObjectDetection}
Current techniques for semantic understanding, including both object detection and semantic segmentation, rely on inputs of 2D images \cite{liu2019edge,ran2018deepdecision}, image and depth data \cite{ophoff2019exploring}, or 3D point clouds \cite{herrmann2021object}. When these data are collected from sensors onboard AR devices, they are frequently subject to distortions due to device motion, occlusions, and resolution limitations (i.e., targets must be observed at an appropriate distance to capture informative data), resulting in incomplete and incorrect knowledge of the environment. Thankfully, just as for spatial understanding, ambient IoT sensors such as cameras and depth sensors can help address these issues by capturing data from additional and more favorable vantage points (e.g., a stationary downward-facing camera). Given that many semantic algorithms are computationally expensive and often best offloaded from AR devices to an edge server, data from IoT sensors can also be transmitted to the edge for processing.

\begin{figure} 
    \centering \includegraphics[width=0.66\textwidth]{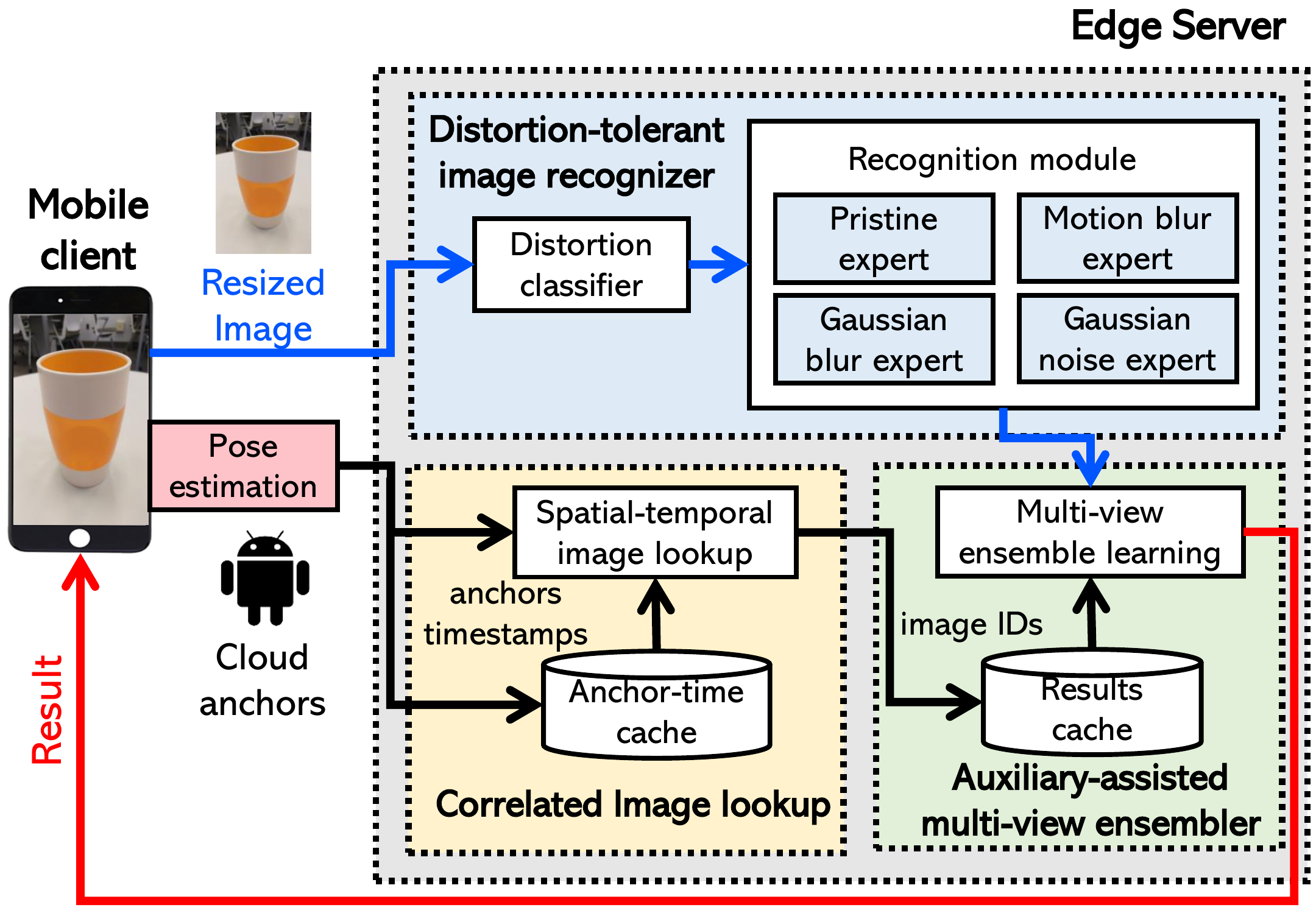}
    \caption{An example of an edge computing-based architecture we have developed for semantic understanding in AR, CollabAR \cite{liu2020collabar}, which provides distortion-tolerant, collaborative image recognition. As an example of combining data from AR devices and IoT sensors to achieve better robustness, we can use high-quality data from IoT cameras to provide the existing viewpoints used in the auxiliary-assisted multi-view ensembler step. 
    }
    \vspace{-0.1in}
\label{fig:CollabARArchitecture}
\end{figure}

One way in which the data sourced from IoT sensors can be used is to combine them with the data from AR devices, to achieve more robust detection. For example, in \cite{liu2020collabar} we developed CollabAR, a collaborative image recognition system in which the camera image from an AR device is combined with spatially and temporally correlated images stored on the edge. The architecture for this system is shown in Figure~\ref{fig:CollabARArchitecture}. Initial inference results based on the device image are provided by the distortion-tolerant image recognizer, then aggregated with the inference results from spatially and temporally correlated images by our auxiliary-assisted multi-view ensembler module, which outputs the final object detection result. This enables CollabAR to achieve over 96\% recognition accuracy for images with severe distortions, with an end-to-end system latency as low as 17.8ms. With ambient IoT cameras, we can quickly provision a large set of high-quality images for the correlated image lookup step, without having to rely on data from AR devices. In general IoT sensors can supplement the data obtained by AR devices to achieve more robust object detection, by providing images or point clouds of the same environment region that are of higher quality, or that contain a more complete view of an object.


Alternatively, the data sourced from IoT sensors may be used to extend the perceptual field of the AR device, to detect objects or surfaces which are outside the field of view or range of device sensors, or objects or surfaces which are entirely occluded. This gives rise to exciting possibilities regarding the extension of the human perceptual field to the entire environment in which an AR user is located. Not only can these IoT sensor data include data from cameras and depth sensors, but one can also incorporate other modalities, towards more reliable recognition that is robust to conditions in which vision-based sensors may perform poorly. For example, passive infrared motion sensors which detect heat can detect human or animal presence in dark environments, while recent works have demonstrated tactile-olfactory-based detection of humans \cite{liu2022star} and chemical sensing of illegal drugs \cite{gao2018superabsorbing} and explosives \cite{li2020qualitative},  which could all be incorporated into ambient IoT sensors. IoT-based monitoring of ambient acoustic signals~\cite{Nelson2018Extending} could be used to improve acoustic-based context detection for audio sources that are located far from the microphones of the AR devices. Central to realizing this vision of semantic understanding through an ambient multimodal sensor network will be solving challenges related to device localization and calibration, variable signal quality, and combining multimodal signals, which we discuss further in Section~\ref{subsec:CombiningDatafromMultipleARUsersandIoTDevices}.\hfill \break

\subsubsection{Enhancing Object State Awareness Using IoT Sensors}
\label{subsubsec:ObjectStateIoTSensors}
In addition to established types of  semantic understanding -- i.e., a knowledge of the type and position of objects present in an environment -- we propose that a knowledge of the current \emph{state} of objects present can also enhance AR applications. This is particularly useful for AR applications that directly interact with physical objects in the surroundings, e.g., overlaying holograms related to an object’s position, orientation, or other properties. These types of AR applications can be enhanced by the understanding of object states, and reflecting changes in real time. While information about some properties (e.g., pose) can be gathered through the processing of images captured by an AR device, IoT sensors incorporated into objects can provide greater robustness to environment properties, as well as data on other types of object states (e.g., strain). Below, we cover several properties of objects that can be obtained through ambient IoT sensors, and discuss potential use cases in AR applications. \hfill \break

\textbf{Pose: }While information about the position and orientation of objects in the surrounding environment can be obtained through marker detection (see Section 4.1.1) or object detection (see Section 4.2.1), these vision-based approaches are often dependent on environmental factors (e.g., lighting conditions or image resolution). Incorporating ambient inertial sensors (e.g., the accelerometer and gyroscope in an IMU) into objects on the other hand provides position and orientation estimates for an object without this dependency on environmental factors. Use cases for integrating inertial sensors into AR for understanding object pose can be seen across various applications that use wearable devices; examples include tracking the orientation of a glove \cite{minh2019motion}, estimating a user’s location and orientation through IMUs in earphones \cite{yang2020ear}, detecting head movements for face-related gestures through smart glasses \cite{masai2020face}, and enabling sensing through haptic devices \cite{shi2020ready, paolocci2020combining}. \hfill \break






\begin{figure} 
    \centering \includegraphics[width=0.85\textwidth]{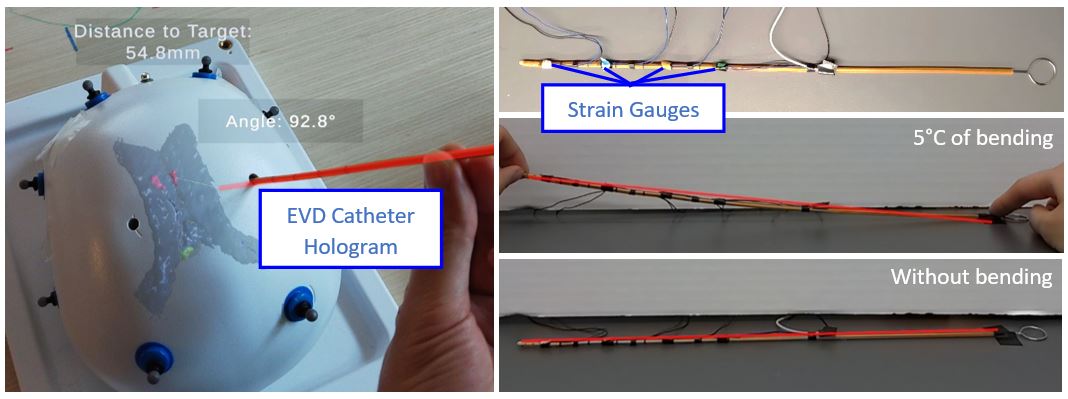}
    \caption{
   The registration of a catheter hologram in an AR-assisted external ventricular drain (EVD) procedure can benefit from the use of IoT sensors for understanding of object state. We used strain gauges to detect the degree of bending on the catheter to align the catheter hologram in AR.}
    \vspace{-0.1in}
\label{fig:ObjectStrain}
\end{figure}

\textbf{Strain: }Strain of an object refers to its deformation due to stress, and can be measured by strain gauges. Strain gauges change the electrical resistance based on the magnitude of the deformation, thus providing knowledge about the deformation of an object such as the bending or external pressure applied. In our recent work, we have been examining this property to enhance the image registration of a catheter hologram in AR-assisted neurosurgery, as shown in Figure \ref{fig:ObjectStrain}. 
We have developed an AR-assisted guidance system for neurosurgery by tracking the position and orientation of the catheter and overlaying a catheter hologram to guide surgeons in targeting the brain ventricle \cite{eom2022ar, eom2022neurolens}. However, due to the deformable shape of the catheter, there were often misalignments between the hologram and the catheter object. Use of fiber Bragg grating (FBG) sensors has been  proposed in multiple lines of work for shape detection in medical applications \cite{lin2018holoneedle, sefati2020surgical, sefati2020data}, 
however the integration of FBG sensors into an AR system is challenging due to its high cost and requirement of an additional measurement device. To address this challenge, 
we are currently experimenting with strain gauges as low-cost IoT sensors that can be used 
to estimate the deformed shape of the catheter. The strain data collected from the catheter are sent to the HoloLens 2 from an edge server to display and align the correct shape of the catheter hologram onto the object. Figure \ref{fig:ObjectStrain} shows the enhancement of catheter hologram alignment by detecting the bending of the catheter using strain gauges. We believe that this will reduce the misguidance that can occur from the misalignment of the catheter hologram in our AR-assisted neurosurgical guidance system, and further enhance the accuracy of catheter placement during the external ventricular drain procedure. \hfill \break

\textbf{Other properties: }Ambient IoT sensors could also be employed to detect other ‘non-spatial’ aspects of object state, as well as pose and strain. This goes beyond simply visualizing the data from existing IoT sensors, to enhancing semantic understanding for AR in new ways. For example, one AR use case is cleaning applications, which inform users if an area requires attention, even if that area does not appear dirty to the human eye. Nanomechanical and electrochemical sensors have been developed which detect pathogens \cite{pujol2020nanomechanical,castle2021electrochemical}, and if integrated into IoT devices, these sensors could provide information about whether objects or surfaces in an environment are contaminated. Similarly, recent works \cite{nguyen2019polydiacetylene,mohammadi2020detection,kim2021microneedle} have shown that food spoilage can be detected using a variety of different devices, from gas, humidity, and temperature sensors to more novel designs using nanomaterials. Deployed in distribution, retail, or culinary environments, these sensors could increase the speed and accuracy of AR-assisted food inventory management, by quickly indicating to workers which items are unsafe for consumption. Extending the perceptive capabilities of users to a wider range of object properties in this manner will provide opportunities to improve both productivity and safety in many industries, and is likely to be a key motivation for the wider adoption of AR.\hfill \break


\subsection{Contextualized Content}
\label{subsec:SensingContextualizedContent}
In the previous two subsections, we explored how ambient IoT sensors may be used to support or enhance two core aspects of next-generation AR, spatial and semantic understanding. We now consider how the virtual content that is presented to the user may be adapted according to the data obtained from IoT sensors. We start by covering \emph{adaptive user interfaces}, the adjustment of virtual content to improve visibility and intelligibility for the user. We then cover the established field of \emph{photometric registration} in AR, the matching of virtual content lighting to real environmental conditions, before extending this to \emph{environment-aware intelligent virtual content}, the provision of virtual avatars or characters which are aware of and respond to a wide range of environmental conditions.

\subsubsection{Adaptive User Interfaces}
\label{subsubsec:AdaptiveUserInterfaces}
The development of adaptive user interfaces, the properties of which are adjusted according to the context they are presented in and the needs of the user, is a long-standing field of research in human-computer interaction (for a recent review see \cite{miraz2021adaptive}). While to a large extent the literature on traditional 2D interfaces has focused on contextual information related to the user (including social and cultural context) or system capabilities, the impact of the diverse and dynamic real environments which will host AR on both human perception of virtual content and system functionality means that environment-aware user interfaces are a vital consideration \cite{lindlbauer2019context, gabbard2007active, macintyre2002estimating}. Below we examine different properties which will inform environment-adaptive user interfaces in AR, and how this will be enabled by ambient IoT sensors.\hfill \break


\textbf{Spatial and semantic understanding: }A number of works on environment-adaptive AR have considered adaptations based on spatial or semantic properties. For example, in \cite{gal2014flare} the authors developed a rule-based framework that enables designers to fit their application components to environments with different geometries, in this case, sourced from the RGB and depth streams of a Microsoft Kinect camera. Similarly, \cite{nuernberger2016snaptoreality} presented a prototype system that allows users to `snap' virtual content to planar surfaces and edges, with environment data again extracted using RGB and depth images from the Kinect. More recently, the work of Lindlbauer et al. \cite{lindlbauer2019context} considers adapting AR user interfaces based on a combination of environment geometry (checking whether a virtual object will be occluded using depth data), user task (which may in part be determined by semantic understanding), and user cognitive load. 

\begin{figure} 
    \centering \includegraphics[width=0.66\textwidth]{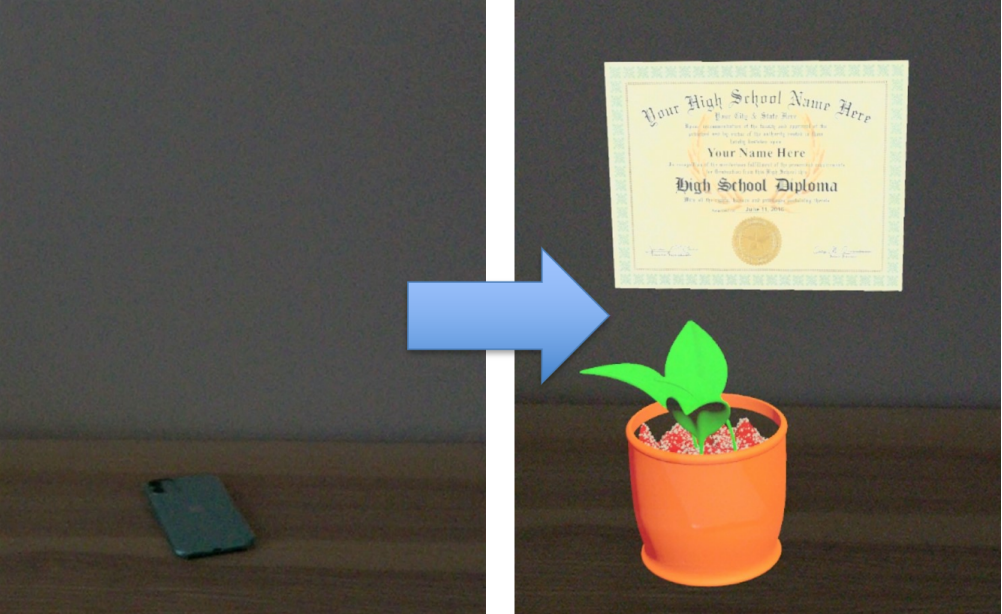}
    \caption{A Magic Leap-based mock-up of an environment-adaptive user interface, which uses semantic understanding of the environment to identify a distracting item (a phone), then covers it with a virtual plant. To further motivate the user to study, a motivational hologram (a diploma) is presented \cite{scargill2022environmental}. 
    }
    \vspace{-0.1in}
\label{fig:AdaptiveUISemantic}
\end{figure}

Regarding semantic understanding specifically, we may wish to place user interface elements near real-world objects that are semantically related, in order to either anticipate a user's current needs, or to place AR-based tasks in context for a smoother experience, as proposed in \cite{cheng2021semanticadapt}. For example, access to AR-guided recipes may appear above the stove, or an in-progress packing list might appear next to a suitcase. Alternatively, we may wish to block the view of real objects with virtual objects -- for instance, in the context of \emph{just-in-time adaptive interventions} (JITAI) \cite{nahum2018just} it may support a user's personal development and change to cover a cigarette packet or mobile phone with another object such as a plant. A mock-up of this on the Magic Leap One headset is shown in Figure~\ref{fig:AdaptiveUISemantic}, and we discuss this topic further in \cite{scargill2022environmental}. Incorporating ambient IoT cameras and depth sensors will enable sufficient levels of spatial and semantic understanding to realize these visions reliably in practical scenarios; accurate, information on the contents of an environment, along with data gathered on how users interact with that environment, will enable the provision of optimally positioned virtual content for AR users. \hfill \break

\textbf{Light, texture, and color: }However, the physical attributes of an environment are not the only factors which should inform the positioning and properties of AR user interface elements -- the light, texture, and colors present also affect visibility and intelligibility. For example, on the OST displays employed on a number of state-of-the-art AR headsets, visibility of virtual content (and as a result usability, user comfort, and sense of presence) is lower at higher levels of illuminance, even those commonly encountered in indoor environments \cite{erickson2020exploring, kim2022investigating, kahl2022influence}. Not only does the additive nature of these displays mean that any virtual content added is less perceptible, but real-world background textures that are visible through dark, transparent regions of virtual content are more visible. Therefore, we may wish to detect when high ambient light levels or highly textured backgrounds are present and increase the brightness (pixel intensity) of virtual content, or adjust the amount of ambient light that is allowed through the headset lenses, to improve content visibility. While state-of-the-art headsets support automatic adjustment of display brightness based on ambient light settings, and the Magic Leap 2 allows users to reduce the amount of ambient light that passes through all or parts of the display \cite{MagicLeapTwoDevice}, the ultimate goal here is automatic, fine-grained adjustment based on both the nature of virtual content and the presence of background textures. 

Furthermore, existing literature has established the effect of contrast ratio and color combinations on content legibility and aesthetics for both traditional (e.g., \cite{lin2003effects}) and AR displays (e.g., \cite{zhan2020augmented}). Given that virtual content may be presented in front of a wide variety of real-world backgrounds in AR, content that is automatically optimized for the current background color and texture is naturally of interest, and was proposed in \cite{gabbard2007active}. On OST displays, blending effects occur in that the perceived color of virtual content is affected by the color of the real-world background \cite{gabbard2020perceptual, gabbard2013color}; the automatic selection of virtual content colors that will either be minimally affected, or affected towards a desired result by blending with the current background, is an important direction for future work.

While light levels, background textures, and colors can be detected through the ambient light sensors or cameras on an AR device, this requires user interfaces (or other content) to be optimized in real-time. This is feasible for the types of backgrounds and content that have been studied so far in AR (the vast majority of existing works consider text readability against backgrounds with a single color or pattern, e.g., \cite{gabbard2007active, debernardis2013text}), but for more complex cases it will likely be advantageous to prepare and provision optimized content in advance. Especially when we consider the complex ways in which light, texture, and color interact to determine visibility, and the potential for combining this information with spatial and semantic understanding of an environment, the use of ambient IoT sensors becomes a necessary addition. To this end, we envision the installation of IoT cameras that capture the texture and color of the real-world surfaces which frequently appear behind virtual content from the user's perspective, with camera poses informed by both analyses of user trajectories and the positions of virtual content. Automatic content optimization, informed by existing work on deep learning for adaptive user interfaces (e.g., \cite{soh2017deep}) can then be performed on the edge, and the result provisioned to an AR user when they enter an environment.\hfill \break

\textbf{Spatial and semantic algorithm performance: }Less considered in the literature is how user interfaces might be adapted based on the current performance of algorithms for spatial and semantic understanding, which is essential to ensure users do not rely on incorrect virtual content. Several pioneering works examined how virtual content may be adjusted in the presence of registration (tracking) error; for example, MacIntyre et al. introduced and applied a \emph{level-of-error} rendering technique \cite{macintyre2000adapting, macintyre2002estimating}, in which a virtual convex hull outlining a real object was expanded according to estimated registration error. This was extended in \cite{coelho2004osgar} to show 2D convex hulls representing possible registration error of virtual objects. In \cite{hallaway2004bridging}, Hallaway et al. demonstrated switching between spatially registered object labels and unregistered augmentations depending on the level of tracking accuracy currently available (this AR system employed a hybrid tracking solution incorporating both ultrasonic and GPS-based tracking, rather than VI-SLAM). Since then, other works have proposed and evaluated visualization techniques to mitigate the effect of registration error \cite{robertson2007using}, as well as alternative ways of visualizing error in navigation tasks (e.g., colored arrows, virtual character expressions) \cite{pankratz2013user}. We are building upon this line of research by exploring new methods of displaying registration error, such as 3D convex hulls around virtual objects. 

\begin{figure*}[htp]
\centering
   \subfloat[AR user interface showing binary classification of tracking performance using symbols and colored planes \cite{scargill2021will}.]{\label{fig:SceneItUIv1}\includegraphics[width=.54\textwidth]{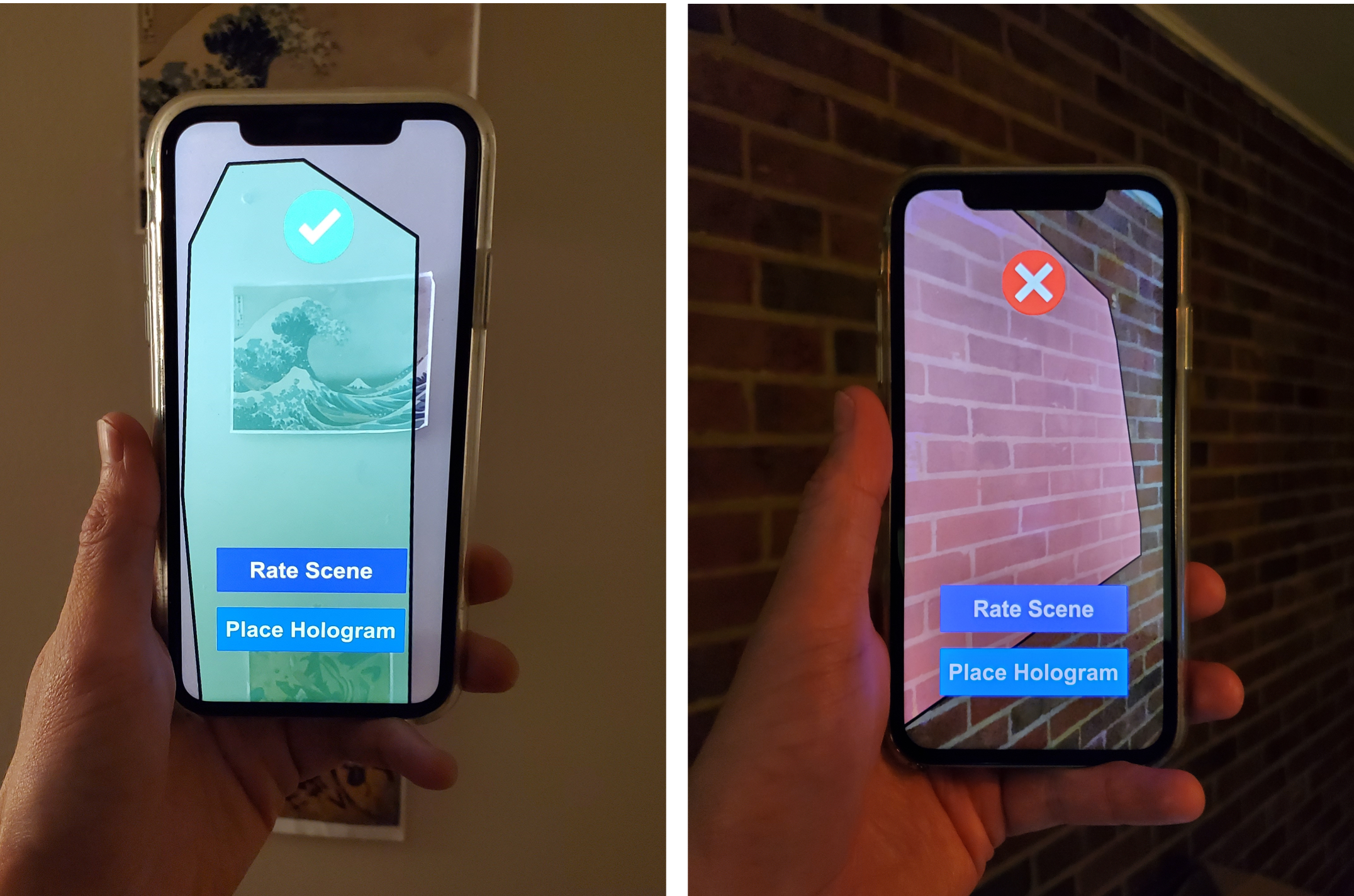}}
   \hspace{\fill}
   \subfloat[AR user interface showing colored binary classification of tracking performance, plus user guidance.]{\label{fig:SceneItUIv2}\includegraphics[width=.37\textwidth]{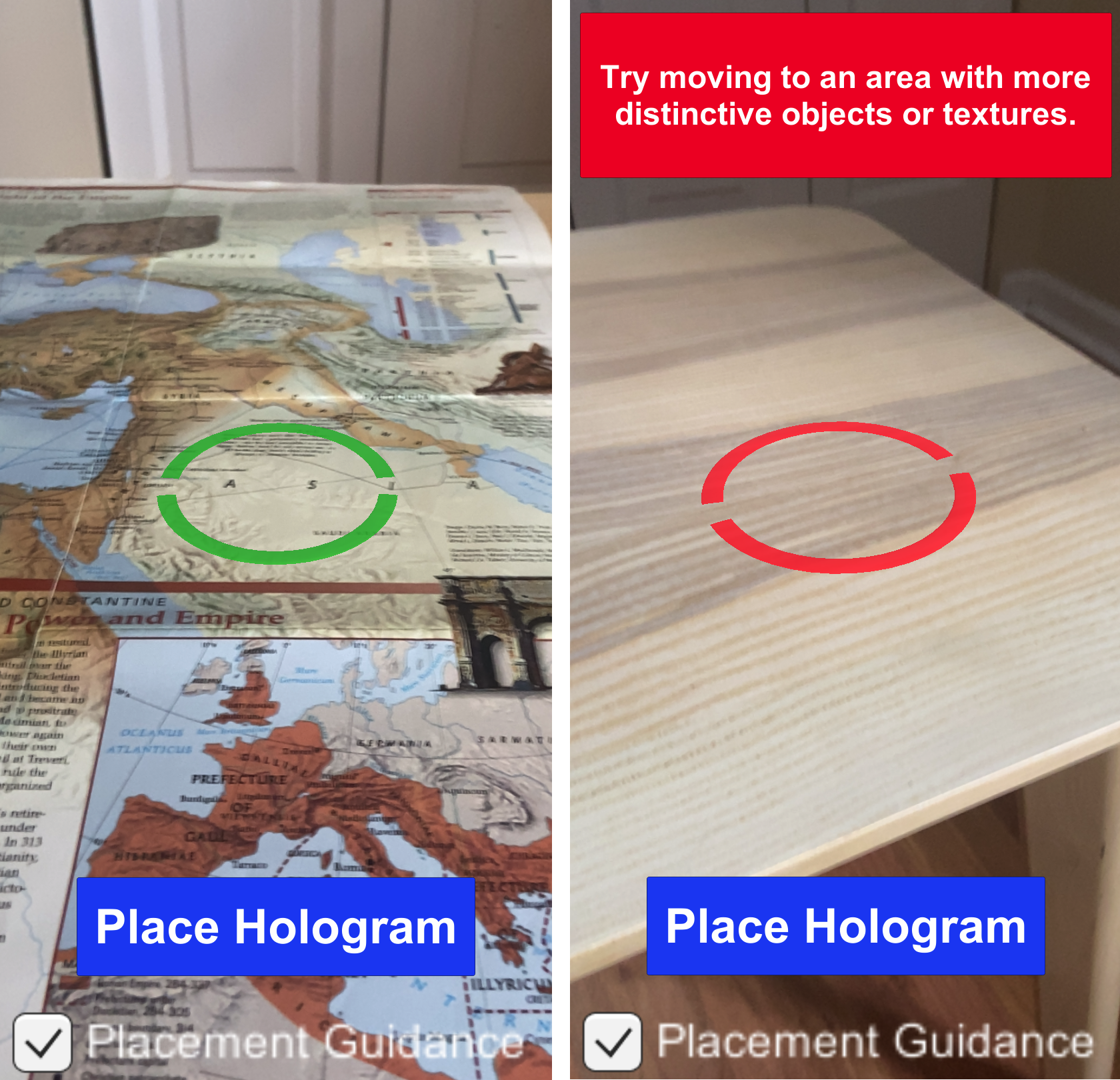}}
   \caption{Two of our prototype adaptive user interfaces for communicating the current level of spatial understanding (device tracking performance) in an environment when placing a virtual object (hologram). Notification of tracking performance classification (a) can be extended to include user guidance on how to improve unsuitable environments (b).}\label{fig:SceneItUI}
\end{figure*}

In particular, our ongoing work examines how to establish and convey the relationship between environmental conditions and tracking or registration error -- as we covered in Sections~\ref{subsec:SensingSpatialUnderstanding} and \ref{subsec:SensingSemanticUnderstanding}, properties such as light and visual texture can impact the performance of the underlying algorithms. Beyond visualizing an estimate of current error, this will enable users and environment designers to take steps to reduce error by altering their environment, before the main AR experience commences. While commercial AR platforms such as ARKit \cite{ARKit} and ARCore \cite{ARCore} indicate when tracking results are unavailable or questionable, along with possible high-level causes, these causes lack granularity in terms of environmental conditions. To address this we developed an interpretable predictive model (a decision tree) for binary classification of tracking performance in \cite{scargill2021will}, along with an example of how environment ratings might be displayed (Figure~\ref{fig:SceneItUIv1}). We have since extended the design of this user interface to use the model output to provide the user with extra guidance, as shown in Figure~\ref{fig:SceneItUIv2}. In these cases the input data for the tracking quality prediction were obtained by the AR device alone, but we envision them being sourced from ambient IoT sensors in the future. We are now developing prediction models which provide more fine-grained estimates of error magnitude, as well as methods to visualize the error estimates associated with different environment regions.  

\subsubsection{Photometric Registration}
Currently, the most established (and well-supported) type of environment-based virtual content adaption in AR is photometric registration -- rendering virtual content such that it is consistent with the lighting in the real environment. Lighting properties of interest include illuminance, light direction, and color temperature, and can be used to more accurately render reflections, specular highlights (the bright regions of a surface that reflect a light source directly), and directional shadows, which in turn increases the level of realism, a user's sense that a virtual object is really present in an environment. Photometric registration for AR has been an active research topic for more than two decades (e.g., \cite{kanbara2004real,gruber2012real}), enabling its recent support in commercial AR platforms such as ARCore and ARKit. State-of-the-art methods in commercial AR (e.g., \cite{ARCoreLighting}) use machine learning to analyze camera images and synthesize environment lighting, similar to recent work on using CNNs for this task (e.g., \cite{mandl2017learning}). Other recent work includes the development of an edge-assisted framework to support real-time light estimation \cite{Zhao2021Xihe}, and the use of physical, geometrically-tracked reflective light probes that are placed in the environment \cite{prakash2019gleam}.   

We envisage ambient IoT cameras and light sensors being employed to enhance photometric registration for two main reasons. Firstly, IoT sensors are able to provide views of the environment not available to AR devices due to occlusions or a limited field of view. This will enable a more complete understanding of lighting within an environment, such as the exact position of light sources which are not visible from the perspective of the AR device. Secondly, unlike AR devices, IoT devices are able to obtain relevant information on environment lighting prior to an AR experience. This means that all the computation related to light estimation does not have to be done in real-time on the AR device; instead, it can be precomputed (and continuously updated) on the edge, both saving resources on the AR device and allowing for more complex light estimation methods to be implemented. Indeed, an example of this type of solution was developed in \cite{rohmer2014interactive}, which the authors termed `distributed illumination', to address the lack of computing power in mobile devices at that time (2014). In this implementation high dynamic range cameras were connected to a stationary PC via a wired connection; we propose extending this to wireless connections using the IoT sensors now available, as well as incorporating more advanced light rendering techniques such as indirect illumination (reflections from the virtual object onto the environment) and the use of spherical harmonics (e.g., \cite{zhao2020pointar}).

\subsubsection{Environment-Aware Intelligent Virtual Content}
Environment awareness in next-generation AR should also extend to the provision of intelligent virtual humans, animals, or other characters that respond naturally to environmental conditions, including the presence and state of real humans or objects, as well as properties such as light and temperature. This has the potential to make virtual content appear even more realistic to users, and further increase a user's sense of presence or immersion. Indeed, recent work has shown that virtual humans whose gaze is directed towards real physical objects supported more effective communication with real participants \cite{andrist2017looking}, and that a virtual human that is able to influence physical objects (e.g., turning off a light) may be seen as more trustworthy by users, and can result in users perceiving a greater degree of social presence \cite{kim2018does}. In \cite{tahara2020retargetable} the authors looked to develop a more sophisticated understanding of semantic context to inform this type of intelligent virtual content, with semantic information first extracted from RGB and depth images, then represented as a 3D scene graph.

While less explored, a virtual character's apparent awareness of and ability to respond to environmental properties such as light, noise, wind, or temperature also holds great potential for increasing levels of realism and immersion. For example, one can imagine a virtual cat that basks in a patch of sunshine, a virtual human that turns its head in the direction of a slammed door, or a virtual character that warms itself next to a heater. In \cite{kim2018blowing} the authors demonstrated the use of IoT wind sensors for AR, with virtual papers that fluttered in response to the airflow generated by a real fan, along with a virtual human that put their hand out to stop the fluttering. IoT sensors will be particularly useful in cases such as this, when detecting or localizing these types of contextual information is beyond the capabilities of the sensors onboard an AR device. Contextual data from different parts of an environment may also be required in advance in order to provision specific animations associated with responsive or intelligent content (e.g., the virtual cat rolling around in a patch of sunshine).

\subsection{Interaction}
\label{subsec:SensingInteraction}
Currently, the primary methods of interaction for user input in AR are tactile, via a touchscreen on the device (e.g., a smartphone), a controller (e.g., the Magic Leap 2), or mid-air gestures via hand tracking (e.g., the Microsoft HoloLens 2, the Varjo XR-3). Some devices also facilitate gaze-based interactions, supported via video-oculography-based eye tracking, or auditory interactions via speech recognition. One important property of all these methods is that their performance can be affected by noise sources in the surrounding environment; for example, the presence of high intensity light sources or high illuminance in general can cause issues for infrared sensor-based depth sensing \cite{tofdesignguide}, resulting in lower quality eye tracking \cite{santini2018pure, scargill2022iot} and hand tracking \cite{guo2022hololens}. 3D gaze point estimation may also be inaccurate at low illuminance due to lower quality spatial mapping \cite{scargill2022iot, guo2022hololens}, while acoustic noise is known to be problematic for speech recognition \cite{zhang2018deep}. Another limitation of tactile and mid-air interaction methods at present is that they only capture gestures made using the hands, and in the case of mid-air gestures captured via hand tracking, gestures are only recognized when the hands are in the field of view of the head-mounted depth sensor. If we compare this to natural human interactions, this is severely restricted; humans frequently express themselves with hand gestures outside this region, as well as gestures with the head, upper and lower body, and facial expressions.

There are two ways in which ambient IoT sensors could be used to manage or address the reduced performance of AR interaction methods due to environmental conditions. One option is to use ambient light sensors or microphones to predict the current level of performance for each interaction method; the most appropriate method could then be suggested to the user, or virtual content could even be adapted to address current limitations (e.g., elements might be increased in size to support gaze-based interactions when gaze estimation accuracy is lower). Another option particularly applicable to speech-based interaction is to use ambient IoT sensors to inform noise cancellation techniques. Such a method was developed in \cite{shen2018mute}, the core idea being that an IoT microphone captures ambient sounds and forwards them to the speaker over its wireless radio, with the information arriving faster than if the sound was captured next to the speaker. This solution is readily applicable to AR audio systems, especially if IoT microphones are already placed in the environment for other purposes. 

There are also a number of possibilities for extending the interaction methods available in AR using IoT sensors. For example, cameras and depth sensors have the potential to greatly expand the range of gestures AR users can use for interaction by capturing a view of a user's entire body. Building upon existing work in human-robot interaction 
(for a review see \cite{liu2018gesture}) and human activity recognition (for a review see \cite{dang2020sensor}), we can develop solutions to recognize natural human gestures as well as relevant activities in AR scenarios. Alternatively, \emph{tangible user interfaces} allow users to manipulate virtual content using a physical proxy, tracked using optical or inertial sensors -- for example in \cite{englmeier2020tangible} a Vive Tracker \cite{ViveTracker} equipped with infrared and inertial sensors was placed inside a physical sphere used for interaction. One could also leverage IoT tactile sensors present in the environment, such as touchscreens, physical buttons, or pressure sensors; a recent example of the latter, particularly relevant to robotic and biomedical applications, was developed in \cite{choi2021tactile}, and is able to detect contact pressure, contact shape, and shear direction. Finally, higher-quality signals available from ambient sensors will also be useful in some circumstances; for example, although the microphones embedded in AR devices are of sufficient quality for speech-based interactions, we may require microphones with a better dynamic range and frequency response for AR applications that present feedback on musical performances to support learning.


\section{IoT-based Actuation for AR}
\label{sec:ARIoTActuation}
Next, we detail how ambient IoT actuators may be used to support or enhance AR experiences. As in Section~\ref{sec:ARIoTSensing}, we examine the different uses which we defined in Section~\ref{subsec:UsesofAmbientIoTforAR}, though given that the role of IoT actuators for AR is less studied than IoT sensors, this section is more exploratory in nature. In this section we combine related uses into one of two subsections; the first covers how IoT actuators can enhance the performance of spatial and semantic understanding algorithms by optimizing environmental conditions, and the second covers how IoT actuators can be used to extend interaction methods or enhance immersion.

\subsection{Spatial and Semantic Understanding}
\label{subsec:ActuationSpatialUnderstanding}
Optimizing environment properties in order to achieve higher quality spatial understanding is an important direction for both marker-based and markerless AR, that will enable both more accurate spatial registration of virtual content and more stable virtual content. Here we cover the use of IoT actuators for marker-based AR in our dynamic marker system (Section~\ref{subsubsec:EnvironmentOptimizationMarkerBased}), and for markerless AR in our environment illuminance optimization system (Section~\ref{subsubsec:EnvironmentOptimizationforMarkerlessAR}). One alternative to these spatial understanding methods was recently introduced in \cite{ahuja2019lightanchors}, and involves the modulation of ambient light sources in a predefined manner; this could be achieved using ambient IoT actuators such as light bulbs, however we focus on established techniques for spatial understanding here. Finally, we also cover the possible use of IoT actuators to optimize environments for semantic understanding  (Section~\ref{subsubsec:EnvironmentOptimizationforSemanticUnderstanding}).

\subsubsection{Environment Optimization for Marker-based AR}
\label{subsubsec:EnvironmentOptimizationMarkerBased}
As introduced in Section~\ref{subsec:BackgroundSpatialUnderstanding}, marker detection is a common technique used for object detection, with the position and orientation of an object obtained through the detection of a printed marker attached to the physical object. To address the challenges of environmental factors (e.g., environment lighting, the distance and angle between an AR device and the marker), we developed a dynamic marker system that controls IoT actuators including an E-Ink display and a smart light bulb to optimize the environment for marker detection. 

\begin{figure} 
    \centering \includegraphics[width=0.90\textwidth]{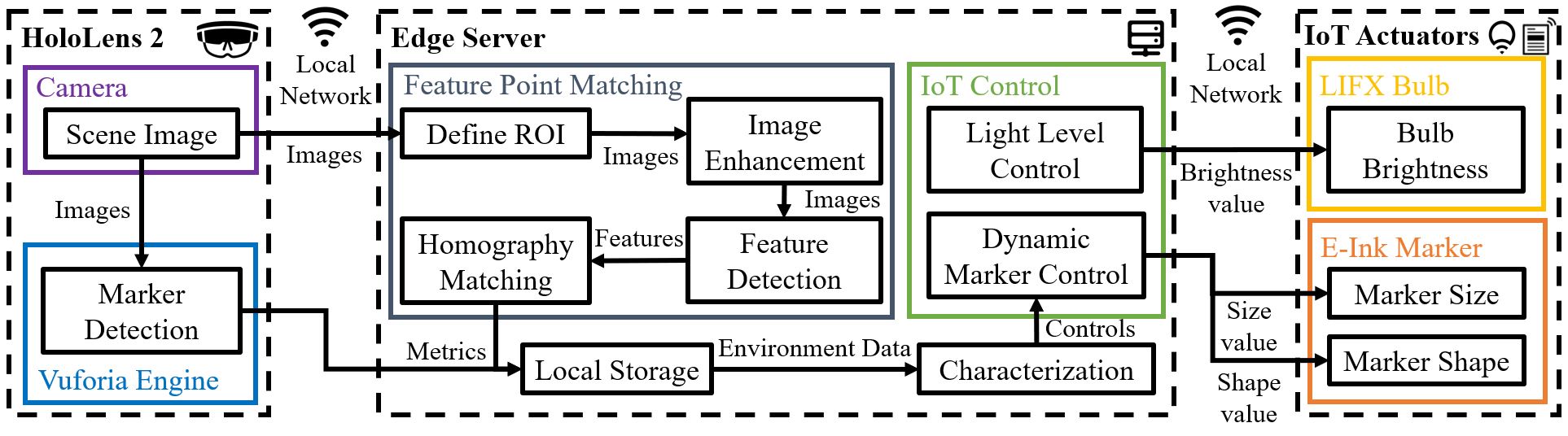}
    \caption{The IoT-based architecture of our dynamic marker system, an example of environment optimization for marker-based AR. Environment images are captured by the AR headset and processed on the edge server for feature detection and matching, to inform environment optimization through IoT actuators.}
    \vspace{-0.1in}
\label{fig:DynamicMarkerArchitecture}
\end{figure}

Dynamic markers were first explored by manipulating the size of projections in a projection-based AR system~\cite{matkovic2005dynamic}, however with recent innovations in low-cost and low-power display hardware (e.g., an E-Ink display), dynamic markers became more practical for many applications. In prior studies, dynamic markers were used in robotic applications such as enhancing human-robot swarm management~\cite{millard2020human}, or aiding Unmanned Aerial Vehicle (UAV) landing by scaling the size of a marker based on the distance to the UAV~\cite{acuna2018dynamic}. Similarly, a dynamic marker has the potential to enhance marker-based AR applications by changing the properties of the marker. Compared to a static printed marker, dynamic markers using an E-Ink display can enhance user interactions by adapting to various situations (e.g., changing the size, shape, or pattern of the marker on the display)~\cite{peiris2011dmarkers}. However, there are challenges associated with a dynamic marker, such as the glare observed on the E-Ink display depending on the environment lighting, which may result in the marker being undetectable by the AR system. Our dynamic marker system for AR further enables the system to control the properties of IoT actuators such as the brightness of a smart light bulb, and the size and shape of a marker shown on an E-Ink display, in order to optimize the environment for marker detection (as shown in Figure~\ref{fig:DynamicMarkerArchitecture}).

\begin{figure}
    \centering \includegraphics[width=0.90\textwidth]{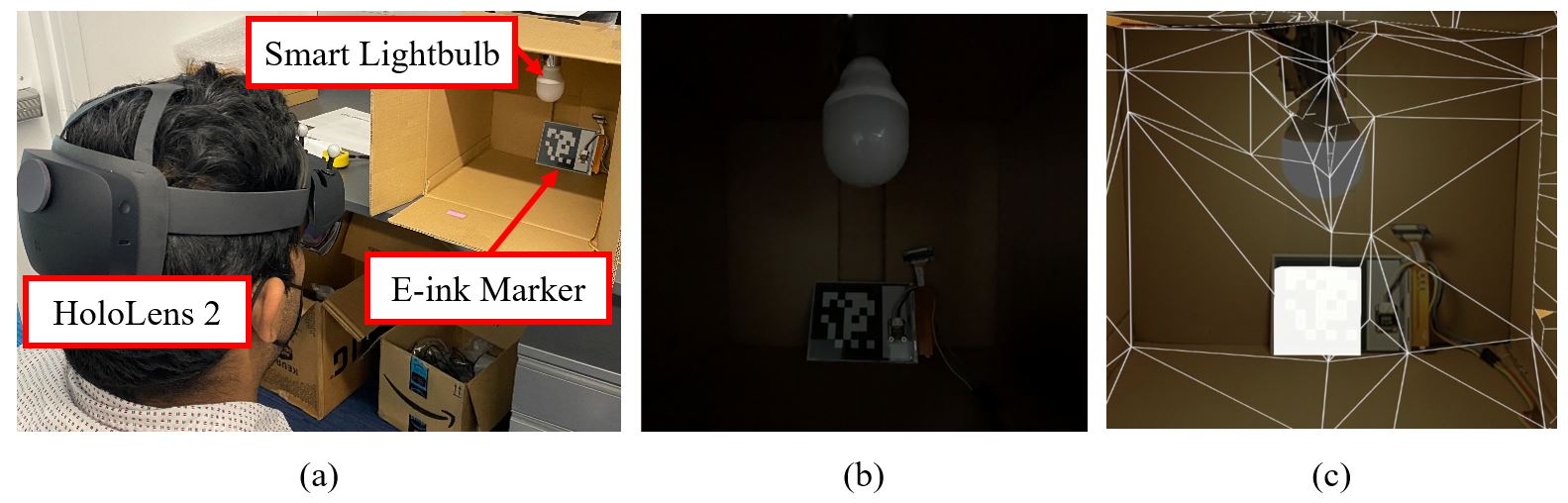}
     \vspace{-0.1in}
    \caption{Hardware setup of our environment optimzation system for marker-based AR: an E-Ink-based dynamic marker, a HoloLens 2, and IoT actuators (a), an undetected dynamic marker in an unoptimized environmental condition (b), an overlay of virtual content through detection of the dynamic marker in an optimized environmental condition (c).}
    \vspace{-0.1in}
\label{fig:EinkHardware}
\end{figure}

The hardware setup of our system (illustrated in Figure~\ref{fig:EinkHardware}) includes a 7.5-inch Waveshare E-Ink display that can show images in 8-bit grayscale with a Raspberry Pi 3, a HoloLens 2 AR headset with Vuforia marker detection, and a LIFX smart light bulb for controlling environment illuminance. As shown in Figure~\ref{fig:DynamicMarkerArchitecture}, we collect scene (environment) images from the HoloLens 2 and process them on an edge server, to characterize the environment with feature points, an important metric for marker detection algorithms, that provides information about the quality of the marker image. Due to the relatively low computational complexity of this task, we use a Raspberry Pi 3 as our edge server. The scene images are first cropped to a region of interest through image processing by detecting the square shape of the marker. We then detect the feature points in the cropped scene image and match them with the reference marker image to quantify the percentage of features available in the scene image. Our system considers three environmental properties (the lighting condition of the scene, the distance from the camera to the marker, and the viewing angle of the camera to the marker) to optimize for marker detection. Environment lighting impacts the quality of the printed marker, with marker detection less robust in darker or brighter conditions. The distance and angle of the camera on the AR headset, determined by the user's movement, impacts marker shape-related factors such as the black and white ratio, edge sharpness, or the information complexity of the marker \cite{kalaitzakis2021fiducial, khan2015factors}. To optimize the environment, we control the IoT actuators to change the brightness of the LIFX bulb, and the size and shape of the marker shown on the E-Ink display, until the percentage of matched feature points reaches the optimal level for marker detection. The latency associated with changing a new image based on the environment characterization in our dynamic marker system takes about one second, due to updating all pixels on the E-Ink display.

To inform this system, we first investigated the relationship between the three aforementioned environment properties and characterization metrics of the scene (i.e., results of feature point matching). We conducted experiments by computing feature matching with four different marker patterns under different environmental conditions. The four markers comprised two fiducial markers in 1-bit grayscale with a similar number of available features from ARToolkit \cite{ARtoolkit} and ArUco \cite{Aruco} open-source libraries, and two image markers; one with a uniform pattern of features and another with an inconsistent pattern of features. We analyzed the changes in the percentage of matched feature points by testing each marker pattern at 9 different viewing distances (20-90cm with an increment of 10cm), 5 different viewing angles (0-60 degrees with an increment of 15 degrees), and 9 different illuminance levels. Each set of environmental conditions was run for 20 trials. Our results showed that there was a correlation between the number of matched feature points and the environment properties. More feature points were matched on the dynamic marker when the viewing distance and angle were lower. On the other hand, fewer feature points were matched on the dynamic marker at lower illuminance levels. We also observed a dissimilitude in the changes in the percentage of the matched feature points among different marker patterns. The matched feature points on the fiducial markers (those from ARToolkit and ArUco) were more dependent on environmental conditions than image-based markers due to their lower number of available feature points. This prompts further investigation to discover the different optimal conditions required for each marker pattern, to achieve more robust and accurate marker detection in AR.

These relationships between environment properties and marker detection performance motivate the use of dynamic markers with IoT-based actuators for various marker-based AR applications. Our dynamic marker system achieves more accurate and robust marker detection by optimizing the environment. For instance, we can change the brightness of a smart light bulb to optimize the level of illuminance, and change the size of the marker shown on the E-Ink display based on the distance from the user to the marker. This environmental optimization enhances user interactions in marker-based AR applications by allowing the system to detect a marker at ease even under initially challenging conditions. Furthermore, a dynamic marker provides much greater flexibility in changing the marker pattern, shape, and size, when compared to a conventional printed marker. This holds potential for marker-based image registration in medical applications by showing different anatomies in AR holograms based on the marker type \cite{eom2022ar, frantz2018augmenting} or in human-robot applications by providing different feedback on the E-Ink display to enable simpler debugging for users \cite{millard2020human}. 

\hfill \break

\subsubsection{Environment Optimization for Markerless AR}
\label{subsubsec:EnvironmentOptimizationforMarkerlessAR}
Given the increasing popularity of markerless AR applications (e.g., \cite{PokemonGo, AmazonARview, IKEAPlace}), and the fact that virtual object instability due to incorrect spatial understanding is still a prevalent issue on state-of-the-art platforms and devices \cite{scargill2021here}, it is also of interest to consider how IoT actuators can be used to optimize environments for better spatial understanding in markerless AR. As we covered in Section~\ref{subsubsec:EstimatingtheQualityofSpatialUnderstandingUsingIoTSensors}, environment illuminance (ambient light level) plays a central role in determining the performance of VI-SLAM, the method underpinning spatial understanding in markerless AR, because it determines the extent to which visual textures are distinguishable for feature-based mapping and tracking. If the tracking cameras on an AR device are pointed towards regions of an environment with sufficiently low or (less commonly in indoor scenarios) high  levels of ambient light, device tracking may fail to initialize, be of lower quality, or be lost, resulting in the incorrect spatial registration of virtual content. Thankfully, illuminance is also readily controlled through IoT actuators which emit light (e.g., smart bulbs) or block light (e.g., smart blinds), leading us to focus our initial efforts on environment illuminance optimization. Additional complexity comes from the need to consider the impact of light on other AR system elements, including the quality of semantic understanding, the performance of eye and hand-tracking algorithms, and the visibility of virtual content on OST displays.

\begin{figure} 
    \centering \includegraphics[width=0.66\textwidth]{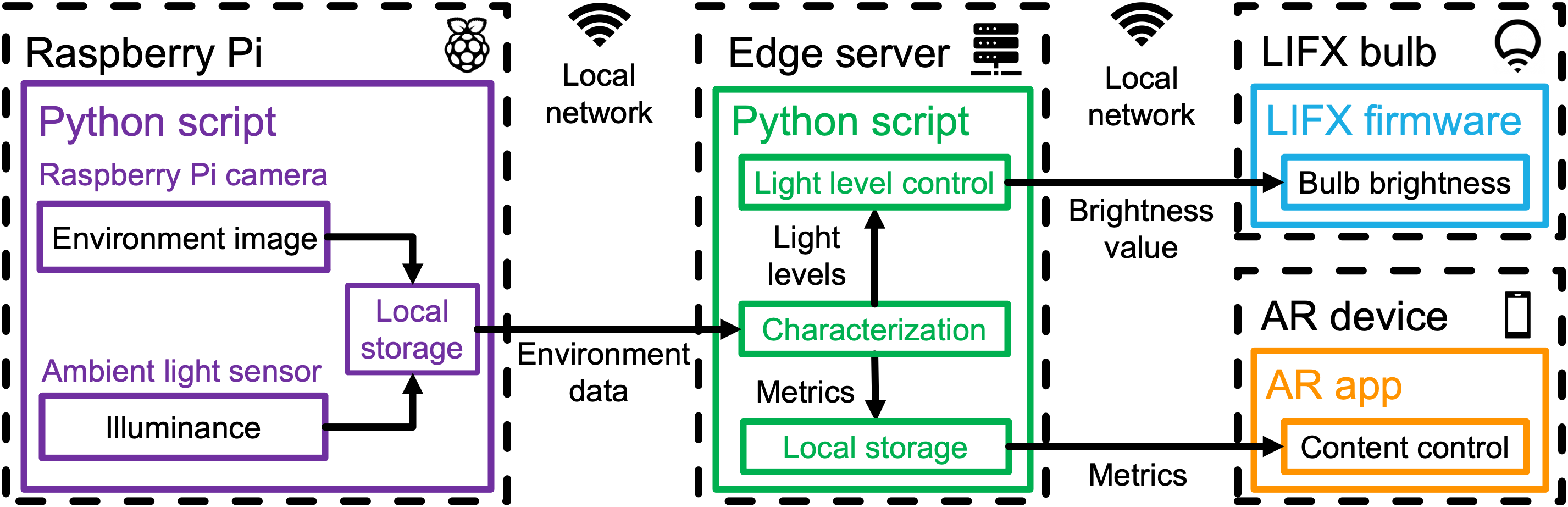}
    \caption{The edge-based architecture for our environment illuminance optimization system for markerless AR \cite{scargill2022iot}. Environment images and illuminance levels are periodically captured by IoT sensors (a Raspberry Pi camera and an ambient light sensor) connected to a Raspberry Pi, which transfers these data to the edge server. Environment characterization is performed on the edge, and instructions are sent to an IoT actuator, a LIFX bulb, to optimize illuminance levels.}
    \vspace{-0.1in}
\label{fig:EnvOptArchitecture}
\end{figure}

In \cite{scargill2022iot}, we developed a proof-of-concept environment illuminance optimization system for AR, which automatically maintains illuminance at a sufficient level for high virtual object stability, and where possible, accurate and precise eye tracking. It uses both IoT sensors to detect current levels of illuminance and visual texture in an environment, and an IoT actuator to control the level of illuminance in that environment. Our results on virtual object position error for different visual textures and illuminance levels (see Section~\ref{subsubsec:EstimatingtheQualityofSpatialUnderstandingUsingIoTSensors}) showed that the robustness of spatial understanding of different illuminance levels is dependent on the properties of the visual textures present, with fine textures requiring greater illuminance to support good performance. Therefore, while our system's default optimum illuminance level is 300 lux (to support accurate eye tracking and low virtual object position error for coarse textures), when the environment contains fine visual textures the core AR functionality, virtual object stability, is prioritized and that optimal level is increased to 750 lux. To offload the computationally expensive environment texture characterization task, while avoiding the transfer of potentially sensitive images to the cloud, we implement an edge-computing architecture, with optimization controlled by a server on the same wireless local area network, as shown in Figure~\ref{fig:EnvOptArchitecture}.

\begin{figure} 
    \centering \includegraphics[width=0.66\textwidth]{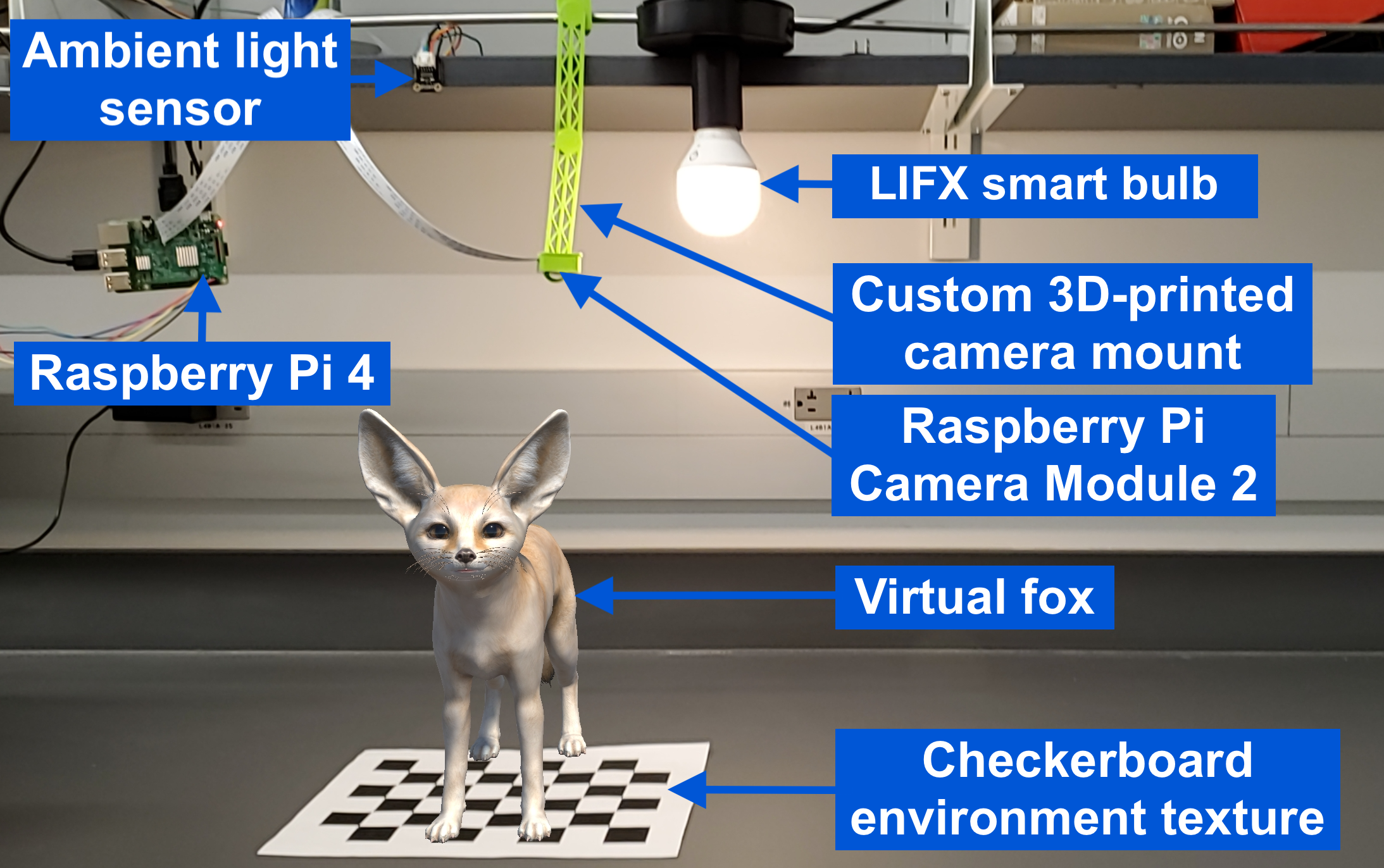}
    \caption{A proof-of-concept installation of our IoT-enabled environment illuminance optimization system \cite{scargill2022iot}, viewed through a custom AR application on a Samsung Galaxy Note 10+ smartphone. In order to capture the appropriate region of the environment with the Raspberry Pi Camera Module 2 we manufactured a custom 3D-printed camera mount, which could be replaced by a programmable mount in future implementations.}
    \vspace{-0.1in}
\label{fig:IoTEnvOpt}
\end{figure}

Our environment illuminance optimization system employs two IoT sensors, an 8MP Raspberry Pi Camera Module 2 and an ambient light sensor (TSL25911FN), connected to a Raspberry Pi 4 (as shown in Figure~\ref{fig:IoTEnvOpt}), to record images of the environment and illuminance every 5s. This environment data is saved to local storage and then sent via an HTTP PUT request to the edge server (a Lenovo ThinkCentre M910T desktop computer). The characterization module determines the optimal light level by detecting the number of FAST corners in the environment image using OpenCV: if more than 250 corners are detected the environment texture is classified as fine, and the optimal light level is raised. If a light level change is required, the characterization module sends the optimal and current light levels to the light level control module. Our system employs one IoT actuator, a LIFX bulb, and the light level control module adjusts the brightness value of the LIFX bulb accordingly, at low latency ($<$0.5s). Environment characterization metrics (illuminance, plus image brightness, contrast, edge strength, and corners) are also stored on the edge for long-term trend analysis, while the latest metrics can be requested by an AR device via an HTTP GET request. The computational and storage requirements of this long-term environment characterization motivated our use of a higher class of edge server (the desktop computer) than the Raspberry Pi 3 used in our optimization system for marker-based AR (Section~\ref{subsubsec:EnvironmentOptimizationMarkerBased}).   

In future work, we will extend our environment illuminance optimization system to employ multiple sets of sensors and smart bulbs to control illuminance in different environment regions. One associated challenge will be device localization and calibration, which we discuss further in Section~\ref{subsec:ARandIoTDeviceLocalizationandCalibration}. It will be desirable for the pose of IoT cameras to be automatically adjustable, such that they capture regions where AR users often point their device camera, or where device tracking is lost. To this end, while our current system uses a fixed camera mount, future implementations could be replaced by a programmable camera mount, controlled according to data gathered on the edge server (the Arducam B0227 Pan Tilt Platform for example provides a low-cost solution for a Raspberry Pi Camera). We will also investigate how other IoT actuators such as smart blinds can be incorporated, as alternative methods of controlling illuminance.

Another direction for future work is to explore how the \emph{visual texture} in an environment could be adjusted to enhance spatial understanding in markerless AR. VI-SLAM algorithms rely on sufficient texture for accurate and robust pose tracking \cite{scargill2022integrated}, but this is often not present in built environments, due to human preferences for minimalism in architecture and interior design. One possible solution would be to detect when an AR experience is taking place in an environment, and use ambient IoT actuators such as electronic or E-Ink displays, or light projectors, to provide additional visual texture during that period. In some cases, additional textures provided by ambient actuators might even be matched with the contents of the AR application, towards greater user immersion (see Section~\ref{subsubsec:ActuationImmersion}). In ongoing work, we are investigating methods that automatically determine where, for a given environment and application scenario, the application of visual texture will be most beneficial. However, any environment texture optimization system must also take into account the impact texture can have on the visual perception of AR users -- greater visual texture can impair visibility on OST displays or distract users from a task \cite{scargill2022integrated}. This is particularly pertinent because some of the environment regions in which applying visual texture is likely to be most beneficial for spatial understanding (i.e., the regions which a device camera is most often facing towards), are frequently the regions where AR content is placed, so adjustment of textures could interfere with human perception. To inform the optimization of visual texture in AR environments, further work is required to investigate the trade-offs between spatial understanding and human visual perception of different visual textures.

\subsubsection{Environment Optimization for Semantic Understanding}
\label{subsubsec:EnvironmentOptimizationforSemanticUnderstanding}
In a similar manner as for spatial understanding, one can also use ambient IoT actuators to optimize environments for semantic understanding. Specifically, IoT light bulbs and blinds can be used to adjust environment illuminance to levels which maximize the performance of vision-based object detection and semantic segmentation algorithms. For example, the guidelines for detecting previously scanned objects on ARKit (described in Section~\ref{subsubsec:BackgroundonObjectDetectioninAR}) recommend an illuminance of 250 to 400 lux \cite{ARKitObjectDetection}. At low levels of illuminance, noise is introduced to the camera images used as input for semantic understanding, especially on mobile devices with a small camera sensor size \cite{anaya2018renoir} -- even small amounts of noise have been shown to result in large accuracy drops for leading image recognition algorithms \cite{liu2020collabar}. Although we covered techniques to obtain images with less distortion using ambient IoT cameras in Section~\ref{subsec:SensingObjectDetection}, one can make this task less challenging by ensuring IoT light bulbs provide sufficient illuminance. On the other hand, high intensity light sources may cause specular reflections on objects or surfaces made of glossy or metallic materials; not only do these artifacts reduce the performance of image-based object detection or semantic segmentation algorithms \cite{anwer2022specseg}, but they also degrade the quality and completeness of data available from time-of-flight depth sensors used to support semantic understanding \cite{tofdesignguide}. IoT light bulbs should be set to intensities that minimize these issues, and IoT blinds should be employed when necessary to block strong sunlight.

Critical to implementing environment optimization systems for semantic understanding will be establishing `regions of interest’, where illuminance-related issues are likely to occur. Regions of interest will be areas of an environment where both (1) illuminance-based distortion or artifacts could occur (e.g., dark corners, or a metallic object close to a light source), and (2) AR users are likely to view, particularly with the expectation of semantic information being provided (e.g., an object of interest). These regions of interest can then be monitored by ambient IoT sensors to inform the control of nearby IoT actuators. For example, if an ambient light sensor on a desk or workbench detects a low light level, the light intensity of a nearby IoT light bulb could be raised. Alternatively, one could detect illuminance-based issues in IoT sensor data directly; for example, specular reflection detection \cite{anwer2022specseg} could be run on the images captured by an IoT camera focused on a metallic object near a window, and when specular reflections are detected the IoT blind on that window would be lowered. We envision these optimization systems being readily combined with the illuminance optimization systems for spatial understanding that we described in Sections~\ref{subsubsec:EnvironmentOptimizationMarkerBased} and \ref{subsubsec:EnvironmentOptimizationforMarkerlessAR}; not only do they share the common goal of minimizing distortion and artifacts in visual data, but regions of interest in need of IoT sensor-based monitoring will frequently overlap.

\subsection{Interaction and Immersion}
\label{subsec:ActuationInteractionImmersion}
As well as being used to optimize the performance of spatial understanding algorithms, IoT actuators may also be leveraged to enhance the user-facing elements of an AR system directly. Below we consider the ways in which a user's interactions with virtual content (Section~\ref{subsubsec:ActuationInteraction}) and their immersion or sense of presence in it (Section~\ref{subsubsec:ActuationImmersion}) may be improved using ambient IoT actuators. 

\subsubsection{Interaction}
\label{subsubsec:ActuationInteraction}
Environmental conditions (e.g., lighting) can impact both the performance of interaction methods such as hand and eye tracking in AR (see Section~\ref{subsec:SensingInteraction}), and an AR user’s perception of visual virtual content (see Section~\ref{subsec:SensingContextualizedContent}), which in turn affects their ability to interact with it. Ideally, we wish to incorporate these constraints into the type of environment optimization systems we introduced in Section~\ref{subsec:ActuationSpatialUnderstanding}, although this will not be without challenges due to conflicting requirements, as we discuss in Section~\ref{subsec:ChallengesEnvironmentOptimization}. In certain cases, for example when conditions cannot be adjusted, or in safety-critical scenarios where the accuracy and speed of task completion are paramount, it may be beneficial to present information on traditional electronic displays that are not as affected by environmental conditions. Users could choose to switch to a nearby ambient visual actuator (e.g., a tablet, smartphone, or smart display such as an Echo Show) to help them complete a specific task, before switching back to the main AR display. Indeed, similar to how screen mirroring is used to share content with a wider audience, this can also be used to support interactions that involve non-AR users. We envision that knowledge of alternative displays that can currently be leveraged will become an important element of environmental awareness in AR.

Another limitation of current AR headset designs is that they facilitate the delivery of visual and auditory sensory information to users, but are not well suited to providing realistic tactile feedback when users interact with AR content. The controllers included with some devices (e.g., the Magic Leap 2) provide uniform tactile feedback and support vibration but do not allow users to actually `touch' AR content, while hand tracking-based interactions allow for natural gestures but do not provide any haptic feedback when users touch virtual content. Recent work has sought to address this latter issue by providing haptic feedback through a vibrotactile ring \cite{meli2018combining,sun2022augmented} or finger-side or nail actuators \cite{preechayasomboon2021haplets,maeda2022fingeret}; however, all solutions involving wearables place extra requirements on user equipment and setup that may not be practical in many scenarios. One potential alternative for visual AR content that is overlaid on a real-world surface is the use of surface haptics, to produce variable tactile sensations and perceived levels of friction on the same surface through electroadhesion \cite{xu2019ultrashiver,mishra2022haptidrag}. By incorporating these devices into an AR environment as ambient IoT actuators, we can program them according to the virtual content that is displayed in a particular region, and thereby provide more realistic haptic feedback to users without the need for wearables.

\subsubsection{Immersion}
\label{subsubsec:ActuationImmersion}
There are a variety of possibilities for using ambient IoT actuators to enhance an AR user’s sense of immersion. One type of enhancement stems from the fact that current AR devices are limited to delivering visual and auditory sensory information; this can lead to a mismatch with other sensory inputs such as the sensation of temperature (thermoception), for example, if a virtual fire is present, but the room is cold. This limits a user's sense of immersion, the feeling that virtual AR content is truly present in the real environment -- research indicates that increasing the modalities of sensory input in virtual environments increases a user's sense of presence in that environment (e.g., \cite{dinh1999evaluating, gooch2009investigation, persky2020olfactory}), consistent with observations that the human perceptual system evolved in multisensory environments \cite{steuer1992defining}. Recent work in VR has explored the potential for different types of external actuators to deliver a range of stimuli specific to the current environment, including speakers for 3D audio \cite{lee2017location}, heat lamps for temperature control, and fans for haptic wind representations \cite{eckstein2019smart} (the authors use the term `smart substitutional reality' to describe their system here), and an olfactometer that provides different scents or smells \cite{archer2022odour}. We propose applying and evaluating these types of systems in AR; for example, if we imagine a therapeutic AR application that relaxes users by placing virtual flowers in their environment, incorporating the scents associated with those plants may increase the application’s effectiveness. Given the variable level of virtual content that may be incorporated into an AR experience, an important consideration will be how a sense of virtual content realism in AR (i.e., a sense that a virtual object is present in a user's physical environment) differs from a user's sense of presence in VR (the feeling that one is actually present in a new virtual environment). 

Alternatively, another type of possible enhancement using IoT actuators involves increasing immersion through the addition of visual and auditory stimuli beyond the current limitations of AR device displays and speakers. In particular, the limited field of view on OST displays and smartphones may be counteracted somewhat by adding visual stimuli, or controlling visual conditions, in the surrounding environment to match AR content. For example, while an AR application displaying the planets of the solar system is being used, the environment light level might be lowered using IoT bulbs, and distant stars represented on the surrounding surfaces using IoT light projectors. Existing immersive art experiences which offer separate projection and VR exhibits (e.g., \cite{VanGoghImmersive}) could provide AR experiences which combine 3D virtual content with 2D projections, while still allowing visitors to interact naturally in the same space. For auditory information, although spatial audio using headphones or the speakers available on some headsets is well-developed, there will be cases (e.g., for a smartphone without headphones, or a headset with a smaller form factor) in which ambient IoT speakers can replicate environmental sound sources much more realistically. Indeed, the opportunities to support smaller AR device form factors by using external devices for virtual content presentation (as well as for computation), in this type of `mixed media AR,' is another important direction for future work.

\section{Challenges and Research Directions}
\label{sec:ChallengesandResearchDirections}
In this section we lay out a set of research directions associated with employing ambient IoT devices to support or enhance AR experiences. First, we describe the use of a game engine-based emulator, required to gather sufficient data on the effect of environment properties to inform the design of IoT-supported AR systems, and prototype them in known, controlled conditions (Section~\ref{subsec:GameEngine-basedEmulationsofIoT-supportedARSystems}). Second, we lay out the challenges associated with AR and IoT device localization and calibration that arise once we implement these systems in the real world, along with potential solutions (Section~\ref{subsec:ARandIoTDeviceLocalizationandCalibration}). We then consider how, given a set of localized and calibrated devices, one might tackle combining the sensor data from those devices (Section~\ref{subsec:CombiningDatafromMultipleARUsersandIoTDevices}). Next, we discuss the challenges of implementing the IoT actuator-based environment optimization systems we described in Section~\ref{sec:ARIoTActuation}, with a focus on the conflicting requirements of different aspects of AR experiences, as well as those that come about once we introduce heterogeneous users and devices (Section~\ref{subsec:ChallengesEnvironmentOptimization}). Finally, we cover security and privacy issues related to platforms which combine AR and ambient IoT devices (Section~\ref{subsec:Privacy}), a vital consideration for the wider acceptance and deployment of ambient IoT-supported AR systems.


\subsection{Game Engine-based Emulations of IoT-supported AR Systems}
\label{subsec:GameEngine-basedEmulationsofIoT-supportedARSystems}
The vision we have laid out in this book chapter 
requires both an in-depth understanding of how environment properties impact AR algorithm performance in diverse scenarios, and the development of deep learning models to predict AR algorithm performance. 
 This is challenging for two 
 reasons. Firstly, we require large amounts of data with accurate ground truth information in a diverse set of environments, which is time-consuming or even infeasible to obtain. For example, obtaining ground truth pose data for VI-SLAM algorithm evaluations in real environments necessitates the use of optical tracking systems such as OptiTrack \cite{OptiTrack} and Vicon \cite{Vicon}, which involves considerable setup and calibration time for each new environment. Secondly, we require fine-grained, systematic manipulation of environment properties in order to perform experiments in repeatable and controlled conditions; this is also difficult to achieve in most real environments because they are often subject to external influences, such as daylight entering through a window, or objects being moved. Indeed, these challenges are why existing datasets for VI-SLAM evaluations (e.g., EuRoC \cite{burri2016euroc}, \mbox{TUM VI \cite{schubert2018tum}}, SenseTime \cite{jinyu2019survey}) or object detection in AR (e.g., \cite{li2020object, ahmadyan2021objectron}) 
 cover a small range of environments or provide no information on environment properties. Instead, we propose the use of highly realistic synthetic data, recently made possible through high-definition rendering techniques in game engines (e.g., Unity~\cite{Unity} and Unreal~\cite{Unreal}) and other rendering software (e.g., V-Ray \cite{VRay}). We developed a methodology for using virtual environments for VI-SLAM evaluations in \cite{scargill2022integrated}, and the process for generating synthetic datasets for semantic understanding was described in \cite{roberts2021hypersim}. We cover the key elements of this approach for ambient IoT-supported AR below.\hfill \break

\begin{figure}
    \centering \includegraphics[width=0.66\textwidth]{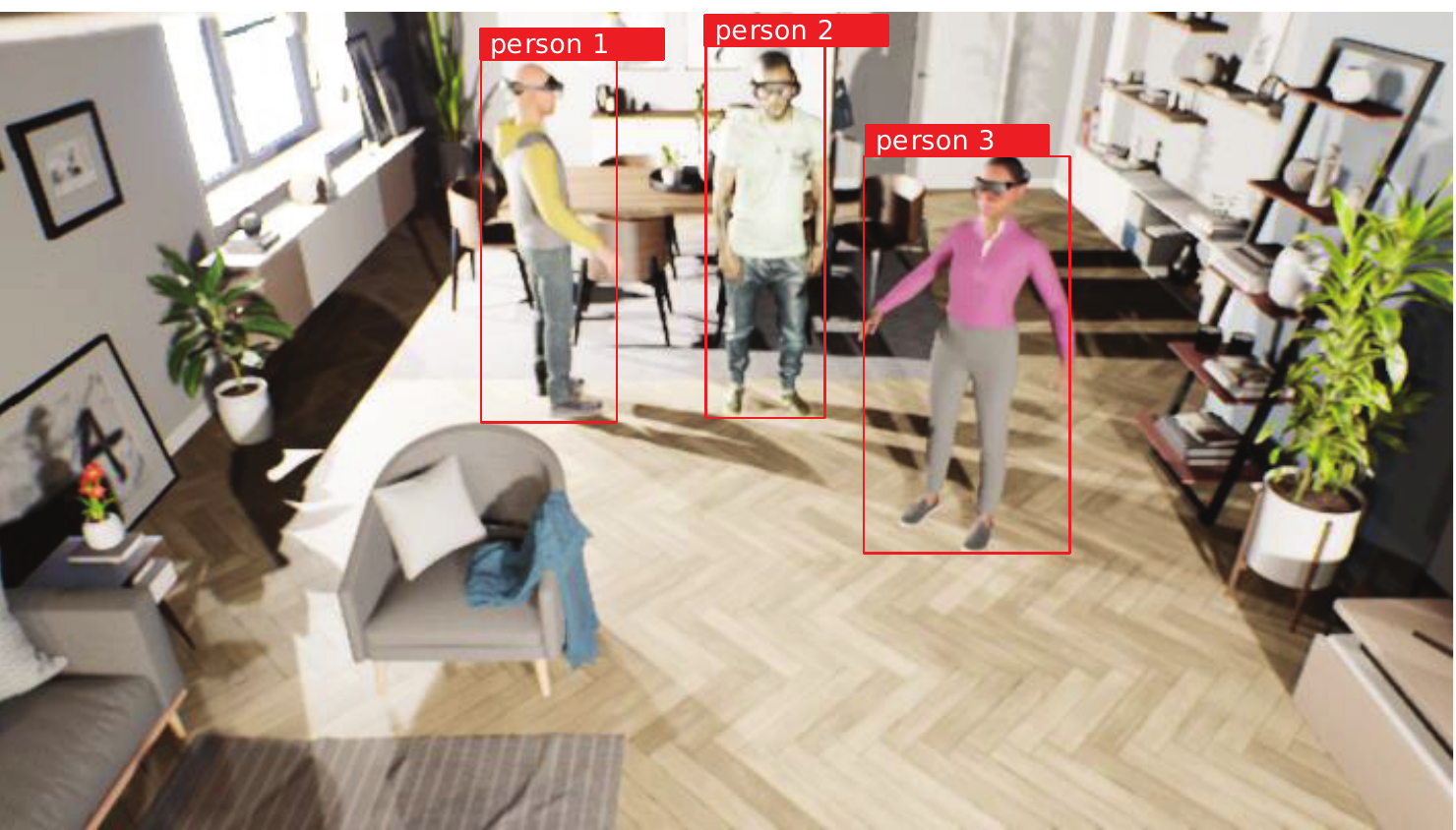}
    \caption{An image captured by an IoT camera in the Unreal-based emulator we have created. The IoT camera overlooks a photorealistic scene with three AR users.
    }
    \vspace{-0.1in}
\label{fig:UnrealSimulation}
\end{figure}

\textbf{AR environment and AR user emulation: }In ongoing work we are using the Unity and Unreal game engines, de-facto standard AR and VR development platforms, to create emulators of AR environments and users. We are using these emulators to generate diverse photorealistic scenes, represent mobile AR user behaviors, and test the performance of a wide range of IoT-supported AR solutions. An example of a scene we have
generated in our initial Unreal-based emulator is shown in Figure~\ref{fig:UnrealSimulation}. Within the generated scenes, we have fine-grained control over multiple parameters related to AR environments (e.g., physical layouts, lighting, reflections, textures), as well as parameters related to AR users and devices (e.g., trajectories, camera properties, motion blur levels). In future we envision the inclusion of other factors to emulate AR environments and users at higher fidelity, such as animating AR users to move and behave like real humans.\hfill \break

\textbf{IoT sensor and actuator emulation: }To enhance the capabilities of our game engine-based emulators, we will also emulate ambient IoT sensors and actuators in AR environments. For example, we will emulate light sensors to measure lighting properties (illuminance, light direction, and color temperature) of the virtual environment. In our preliminary investigations we have added an IoT camera in our emulator: Figure~\ref{fig:UnrealSimulation} shows an image captured by this IoT camera, overlooking a space with three AR users. The camera has also identified and tracked the AR users in its view. We are currently exploiting the emulation of IoT cameras to develop algorithms that increase the accuracy of spatial and semantic understanding, by fusing the images from these IoT cameras with the sensor data captured by the AR devices worn by emulated AR users. 
Adding the emulation of IoT actuators (e.g., by creating and modifying light sources) will enable the modeling and prototyping of environment optimization systems, such as those we described in Section~\ref{subsec:ActuationSpatialUnderstanding}. \hfill \break

\begin{figure} 
    \centering \includegraphics[width=0.98\textwidth]{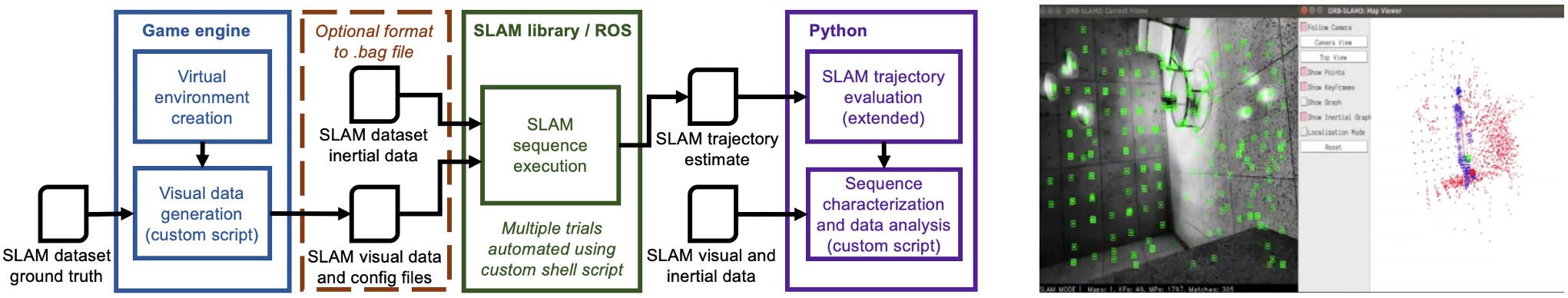}
    \caption{An overview of our methodology (left) and a screenshot of the SLAM sequence execution step (right) for our game engine-based emulator for VI-SLAM using virtual environments~\cite{scargill2022integrated}.}
    \vspace{-0.1in}
\label{fig:VirtualInertialSLAM}
\end{figure}

\textbf{Ground truth generation: }As well as controlling parameters related to environmental conditions, we envision building pipelines that automatically generate the ground truth data required for algorithm evaluation. We obtained the ground truth of the states (e.g., camera poses, sensor positions) of AR devices and IoT sensors in our prior work~\cite{scargill2022integrated}, and will use the pipelines to generate
pixel-wise semantic instance annotations for images captured by AR devices and IoT sensors. For example, in \cite{scargill2022integrated} our VI-SLAM evaluation methodology used existing SLAM dataset ground truth trajectories to generate sequences in new virtual environments, while preserving the use of real inertial data, as shown in Figure~\ref{fig:VirtualInertialSLAM}. In our future work, we envision exploiting the automatic generation of ground truth data, especially fine-grained pixel-wise data, to evaluate IoT-supported AR systems under diverse environment settings, such as assessing mapping accuracy in SLAM, which is otherwise difficult to evaluate in real environments~\cite{mappingground}.

\subsection{AR and IoT Device Localization and Calibration}
\label{subsec:ARandIoTDeviceLocalizationandCalibration}
In the game engine-based emulators described above we have access to the ground truth poses of AR and IoT devices, along with the parameters of IoT sensors and actuators (e.g., correct information on environment regions captured by sensors and affected by actuators such as light sources). However, a critical challenge for multi-device, IoT-supported AR in real environments is how we establish all AR and IoT devices’ poses within the same frame of reference, and determine the environment regions either in view of sensors or which will be adjusted by an actuator. Existing literature provides us with potential ways to address this problem; for example, the localization of IoT devices is a known issue, and a variety of different solutions have been proposed, including the use of ultra-wideband signals \cite{beuchat2019enabling} and  retroreflectors \cite{shao2018retro} -- for a survey of indoor localization systems and technologies see \cite{zafari2019survey}, and for discussion of the challenges specific to IoT devices, see \cite{ghorpade2021survey}. Established camera calibration techniques are also an option for IoT cameras, such as the OCamCalib toolbox \cite{OCamCalib} employed in \cite{rohmer2014interactive} for photometric registration using multiple cameras. For localizing light sources a possible approach is to build on the work in \cite{gardner2017learning}, and apply DNNs to images obtained by AR devices or IoT cameras. Methods are also available to localize multiple AR devices; cross-platform solutions include Azure Spatial Anchors \cite{AzureSpatialAnchors} and Google Cloud Anchors \cite{GoogleCloudAnchors}, while the `ARWorldMap' feature can be used on ARKit \cite{ARKitARWorldMap}, and `Shared Spaces' on the Magic Leap \cite{MagicLeapSharedSpaces}. The above techniques provide both inspiration and readily implementable solutions for systems which employ AR and IoT devices.  

Based on our analysis of existing techniques, we identify a number of possible approaches for localizing and calibrating both AR and IoT devices, and consider their advantages and disadvantages. The solution that is perhaps most readily implemented for current commercial AR devices is to identify IoT devices within the frame of reference established by an AR platform. For example, the pose of devices could be recognized by displaying markers on them or attaching printed markers to them (see Section~\ref{subsec:BackgroundSpatialUnderstanding} for background on spatial understanding using markers), although this may not be feasible for many small devices, and has downsides in terms of environment aesthetics. Alternatively, spatial anchors could be manually aligned with IoT devices~\cite{SpatialAnchors}; however, this approach could be both labor-intensive and time-consuming, and the accuracy that can be achieved requires further investigation, as it may be susceptible to human error or easily impacted by environmental factors. On the other hand, a different research direction is to identify AR device poses in the views obtained by IoT cameras, aided by motion and visual cues from the AR device  -- this may be particularly useful for collaborative SLAM techniques (e.g., to inform loop closing). For the mapping of available environment control actions (e.g., increasing the brightness of a light bulb or lowering a smart blind) to environment properties as detected by an IoT sensor or AR device, we plan to develop efficient sampling strategies for long-term monitoring of environment properties, such that we can automatically determine the necessary control actions to achieve specific conditions, and thereby optimal AR experiences.    



\subsection{Combining Data from Multiple AR Users and IoT Sensors}
\label{subsec:CombiningDatafromMultipleARUsersandIoTDevices}
Along with localizing and calibrating AR devices, IoT sensors and actuators, 
we must also consider how to combine the data from the variety of different sensors that are available in the environments that host AR experiences. This sensor data is diverse, including multimodal sensor data of AR devices and IoT sensors captured from different vantage points. Below we discuss three challenges related to combining data from multiple AR users and IoT sensors, and outline the associated research directions. \hfill \break

\textbf{Communication-efficient AR 
with multiple users and multiple IoT sensors:}
In a multi-user, multi-sensor AR scenario, as the number of AR users and IoT sensors grows, the operational overhead of the IoT-supported system such as bandwidth consumption also scales~\cite{ran2020multi},  due to the transmission of large amounts of contextual data from multiple users or their ambient environments.  The system also needs to maintain a low user-perceived latency~\cite{apicharttrisorn2020characterization} to ensure seamless integration with the real world and synchronous perception of contextual information among different users.
To improve communication
efficiency and reduce user-perceived latency, we envision  developing intelligent network-adaptive AR-IoT solutions that adapt the size and frequency of data transmissions  under changing network conditions. Since multiple users and IoT sensors that are in close proximity usually exhibit a strong correlation in the perceived contextual information, 
we also envision exploiting the collaboration among multiple AR users and IoT sensors for communication
redundancy elimination.
\hfill \break

\textbf{Assessment and processing of data with different signal quality levels: }
Another research direction in this space is the quality assessment of data from different devices and the development of robust approaches for combining this data. 
The signal quality levels obtained from different vantage points will change over time due to a variety of factors; for example, other noise sources will be interfering with signals captured by different microphones, users of AR devices will move closer and farther from different IoT-based cameras, and IoT sensors in unusual poses will cause the misbehavior of the object detection models that perform well for IoT sensors in canonical poses. We envision the real-time and in-situ evaluation of the quality levels of data captured by different devices at different times.
After the data quality assessment, we will seek to combine signals with different quality levels; we envision designing algorithms that will adaptively amplify `good' signals and place less weight on signals that are less useful, e.g., by integrating the existing design of appropriate attention mechanisms~\cite{vaswani2017attention}.\hfill \break

\textbf{Multimodal data: }In IoT-supported AR systems, AR devices and IoT sensors continuously capture and process 
multimodal data (e.g., visual, auditory, haptic, olfactory, and thermal sensory information). One potential research direction is multimodal sensing and learning in which the visual modality enhances, or is enhanced by, other sensing modalities; another is fusing sensor data with different temporal and spatial traits. Given the heterogeneity of the collected data, we envision capturing correspondences and transferring knowledge between modalities to improve the performance of various aspects of AR functionality, including using gesture and speech for precise multimodal interaction, and using thermal, light, and visual sensors for detecting scene changes throughout the day.
\hfill \break



\subsection{Optimizing Environments with Conflicting Requirements}
\label{subsec:ChallengesEnvironmentOptimization}
Environment optimization systems which automatically adjust properties such as light and visual texture (see Section~\ref{subsec:ActuationSpatialUnderstanding}) have the potential to facilitate AR experiences of both higher and more consistent quality. However, before these types of systems are ready for deployment, consideration must be given to how we manage conflicting environmental requirements. These conflicts, already a challenge for designers of spaces which host AR, will be commonplace in practical scenarios, as we describe below.  \hfill \break 

\textbf{Conflicting environmental requirements: }Even in an environment in which just one AR user is present, adjusting environmental conditions to optimize one aspect of an AR experience may have negative effects on another element of system functionality or the overall user experience. For example, the addition of visual texture to an environment may improve spatial understanding, but be distracting for the user \cite{scargill2022integrated}. For headsets with an OST display, the low level of illuminance which maximizes content visibility may result in a reduced level of performance for spatial understanding or eye-tracking algorithms \cite{scargill2022iot}. Furthermore, environmental requirements related to AR may also be at odds with constraints related to energy efficiency or the comfort of human occupants in general. Any environment optimization system will need to (1)~define the characteristics of the AR devices, applications, and users to be served, in order to define a reasonable `operating range' in which the system is functional and usable; (2)~apply other environmental constraints, e.g., those imposed by building managers; (3)~implement real-time analysis of the system elements and environment regions users are interacting with, in order to determine and prioritize environment adjustments. This latter element is a particularly complex challenge, necessitating innovative solutions to manage continuously updating environment maps and device poses, and to avoid oscillations between environment states that may occur due to conflicting adjustments. Some of these issues were recently tackled in the context of shared control of public IoT actuators \cite{kim2021hivemind}, including device discovery, establishing `areas of interest' for IoT devices, and aggregating conflicting requirements; this work may serve as a valuable starting point for developing AR-specific systems.\hfill \break

\textbf{Heterogeneous AR users and  devices: }The environment optimization challenge becomes even more complex when we consider the differing requirements of heterogeneous AR devices and users. Different devices are equipped with different sensors, run different marker detection or VI-SLAM algorithms, and employ different types of displays: smartphones may require greater levels of illuminance and texture to achieve acceptable tracking performance \cite{scargill2021here}, while OST displays may require lower illuminance than VST displays to achieve the same level of content visibility~\cite{Giorgio18Perception}. Heterogeneous AR users will have different behavior patterns (e.g., mobility characteristics, eye gaze patterns, virtual content interactions), as well as different expectations and requirements for the properties of virtual content. For example, school children exploring an AR science exhibit may move rapidly, prompting the addition of environmental textures to maintain high-quality pose tracking. On the other hand, an academic carefully examining virtual content in the same environment might require a view that is uninhibited by a textured background, and also be more concerned with the stability of virtual content. While knowledge of the type of AR devices present in an environment is likely to be relatively straightforward in most cases, capturing and analyzing information on the properties of AR users is a more complex task, which will require ongoing analysis of both individual users and user populations. This in turn will have significant security and privacy implications, which we discuss next in Section~\ref{subsec:Privacy}.

\subsection{Secure and Privacy-Preserving AR-IoT Platforms}
\label{subsec:Privacy}
The development of secure and privacy-preserving AR-IoT platforms is paramount for achieving widespread societal acceptance of AR systems incorporated with ambient intelligence technology.  
\hfill \break

\textbf{Security:} 
The increase in the number of connected AR and IoT devices and the richness of data collected by them bring security concerns to AR-IoT platforms. 
Building upon the categorization adopted in~\cite{lebeck2018towards,roesner2014security}, we classify concerns as related to  security 
of AR devices~\cite{ruth2019secure,dechicchis2019adaptive,lebeck2018towards,roesner2014security} and IoT sensors and actuators~\cite{neshenko2019demystifying,antonakakis2017understanding} as related to \mbox{\emph{input security}} or \mbox{\emph{output security}}. 
Input security challenges are well known in traditional arenas of networked and computing systems, but become even more important for AR-IoT platforms with rich multimodal inputs (e.g., RGB, depth, inertial, haptic, auditory sensor readings, and even gaze tracking data) of AR and IoT devices. 
Compromised inputs will degrade the performance of AR-IoT platforms, including localization and tracking accuracy, rendering quality, and the accuracy and completeness of environmental awareness. To mitigate input security risks, potential research directions include designing input validation and sanitization methods that apply to multimodal data from AR and IoT devices, and building
 fault-tolerant AR-IoT platforms robust to corrupted input data. 
 AR outputs 
 produced by malicious or bug-ridden applications can also be potentially harmful or distracting. For example, in the case of industrial AR in a smart warehouse, it would be dangerous for a visual overlay to obstruct the operator's view; IoT actuators that generate flashing visuals, shrill sounds, or intense haptic signals could cause physiological damage to the user. To mitigate output security risks, inspired by existing works on adaptive policies to secure visual outputs in AR devices~\cite{ahn2018Adaptive,dechicchis2019adaptive}, we envision the development of reinforcement learning-based policies that prevent distraction due to tampered content of AR devices or IoT actuators, with these policies able to adapt to dynamic environments through trial-and-error.
\hfill \break

\textbf{Privacy: }AR-IoT systems'
need for rich, continuous sensor data (from both AR and IoT devices) raises privacy concerns for both users and
bystanders in the AR environment. 
Buggy or malicious applications may record privacy-sensitive information on  user states or the surrounding AR environment by compromising the AR devices~\cite{yang2022study} or IoT sensors~\cite{tawalbeh2020iot}. To mitigate privacy risks, we envision enforcing privacy-preserving policies to limit access to
potentially sensitive sensor data. For example, we will anonymize
 human faces with an extremely low resolution before feeding images to object detection applications running on AR or IoT devices; we will seek to prevent applications from inferring sensitive information about an AR environment (e.g., the contents of a home) by concealing information contained in the point cloud constructed by VI-SLAM algorithms.


\section{Conclusion}
\label{sec:Conclusion}
In this book chapter, we explored the variety of ways in which ambient IoT sensors and actuators could be used to support next-generation AR. We categorized these uses by defining five different aspects of AR system functionality or user experience that may be enhanced -- spatial understanding, semantic understanding, contextualized content, interaction, and immersion -- and provided an overview of relevant IoT devices (Section~\ref{sec:AmbientIoTforAR}). We then examined the possibilities for each of these categories in detail, when employing IoT sensors (Section~\ref{sec:ARIoTSensing}) and actuators (Section~\ref{sec:ARIoTActuation}). Finally, we discussed a number of research directions related to implementing ambient IoT-supported AR systems, along with associated challenges and opportunities for future work (Section~\ref{sec:ChallengesandResearchDirections}).

One important point to make, having reviewed multiple different uses of IoT sensors and actuators, is that the devices physically deployed in a given environment will likely be used for multiple purposes concurrently (e.g., images from a single IoT camera may serve as input to spatial understanding, semantic understanding, photometric registration, and visual texture estimation algorithms), so consideration will need to be given to balancing the needs of each use in these cases. Similarly, our vision for ambient IoT for AR in general, is that it is combined with the use of ambient IoT devices for other purposes; for example, IoT-based systems incorporating devices also applicable to AR have been proposed for energy efficiency \cite{salman2016energy}, occupant comfort and productivity \cite{aryal2018smart}, surveillance \cite{zhang2015design} and building ventilation \cite{chhaglani2022flowsense}, while in \cite{tractinsky2012considering} the authors consider the role of ubiquitous displays in environment aesthetics. Moreover, as AR becomes more and more integrated into our lives in the coming years, we envision that built environments, products and materials will be designed with ambient intelligence for AR in mind, in order to support the use cases we have laid out in this book chapter.

\section{Acknowledgements} 
We thank Guohao Lan, Zida Liu, Yunfan Zhang, Jovan Stojkovic, Achilles Dabrowski, Alex Xu, Ritvik Janamsetty, Tiffany Ma, Owen Gibson, Michael Glushakov and Joseph DeChicchis for their contributions to this work. This work was supported in part by NSF CAREER Award IIS-2046072, NSF grants CNS-2112562 and CNS-1908051, Facebook Research Award, IBM Research Award, and a Thomas Lord Educational Innovation Grant.

\bibliographystyle{abbrv-doi}
\bibliography{references}

\end{document}